\documentclass[11pt,letterpaper]{article}
\usepackage[T1]{fontenc}
\usepackage{lmodern}
\usepackage[margin=0.95in]{geometry}
\usepackage{amsmath,amssymb,amsthm,graphicx,array,multirow,booktabs}
\usepackage{caption}
\usepackage[numbers,sort&compress]{natbib}
\usepackage{tikz}
\usetikzlibrary{patterns,decorations.pathreplacing,shadows,shapes.geometric}
\usepackage[hidelinks]{hyperref}

\theoremstyle{definition}
\newtheorem{theorem}{Theorem}
\newtheorem{lemma}[theorem]{Lemma}
\newtheorem{proposition}[theorem]{Proposition}
\newtheorem{corollary}[theorem]{Corollary}

\newsavebox{\tablebody}
\newcommand{\F}{\mathbb{F}}
\newcommand{\wt}{\operatorname{wt_s}}

\newcommand{\cmins}{c_{\min}}
\newcommand{\ea}[4]{[\![#1,#2,#3;#4]\!]}
\newcommand{\eaa}[4]{(\!(#1,#2,#3;#4)\!)}
\newcommand{\Cc}{\mathcal{C}}
\newcommand{\Cd}{\mathcal{C}^{\perp_{\mathrm{s}}}}
\newcommand{\Hull}{\mathrm{Hull}(\mathcal{C})}

\definecolor{achieve}{RGB}{31,119,120}
\definecolor{refute}{RGB}{178,58,50}
\definecolor{bdh}{RGB}{110,110,118}
\definecolor{ghw}{RGB}{62,102,158}

\begin{document}
\raggedbottom
\title{Validity and maximum violation of the entanglement-assisted quantum Singleton bound for linear stabilizer codes}

\author{Jae-Hyun Baek$^{1,2}$\thanks{These authors contributed equally to this work.}
\quad Byung-Sun Won$^{3}$\footnotemark[1]
\quad Jon-Lark Kim$^{4,2}$\thanks{Contact author: \href{mailto:jlkim@sogang.ac.kr}{jlkim@sogang.ac.kr}.}\\[0.8em]
\small $^1$Department of Artificial Intelligence, Sogang University, Seoul, Korea\\
\small $^2$Institute for Mathematical and Data Sciences, Sogang University, Seoul, Korea\\
\small $^3$Ewha Institute of Mathematical Sciences, Ewha Womans University, Seoul, Korea\\
\small $^4$Department of Mathematics, Sogang University, Seoul, Korea}
\date{}
\maketitle

\begin{abstract}
Entanglement-assisted quantum codes protect quantum information with
entangled pairs shared in advance, and the entanglement-assisted quantum
Singleton bound of Brun, Devetak and Hsieh is the standard estimate of how
many pairs they consume. Codes that violate it at large distance are known.
For linear entanglement-assisted stabilizer codes, which include all qubit
stabilizer codes of this kind, we determine when the bound holds and how
far it can fail. The answer depends on how many
qudits, together with the receiver's halves of the pairs, recover the
logical qudits. If this number is less than twice the number of logical
qudits, the bound holds, also for degenerate codes. If it is exactly twice
that, the bound overstates the number of pairs by at most the length of
the longest classical maximum distance separable (MDS) code of that
dimension, minus the dimension. Codes carrying values and derivatives of
polynomials on the projective line attain this maximum whenever the field
size is at least that dimension and the MDS conjecture is known for it,
which includes every prime and, by a recently announced result, every even
field size. If the number is more than twice the number of logical qudits,
the violation is bounded in terms of the lengths of classical codes of a
given Singleton defect, and the bound is attained for one logical qudit.
Apart from cited MDS lengths, the proofs are formalized in the
Lean proof assistant.
\end{abstract}

\section{Introduction}

In entanglement-assisted quantum error-correcting codes
(EAQECCs)~\cite{BDH2006,WildeBrun2008}, a sender and a receiver share $c$
maximally entangled pairs in advance and use them to protect $k$ logical
qudits sent through $n$ uses of a noisy channel. With entanglement available, any
classical linear code can be turned into a quantum code, and the resulting
parameters can exceed what is possible without assistance. The pairs must
be distributed and stored, so a basic question, asked by Luo, Ezerman,
Grassl and Ling~\cite{Luo2022}, is how much entanglement a code of given
length, dimension and distance needs. Constructions give upper bounds
on this amount, and most of the literature is about
constructions~\cite{Luo2022,Galindo2019,Li2026}. Lower bounds require
impossibility proofs. We study them for EA stabilizer codes whose
stabilizer is $\F_q$-linear, written $\ea{n}{k}{d}{c}_q$ with qudits of
dimension $q$. The standard constructions from classical linear codes give
such codes, and so do Grassl's teleportation codes~\cite{Grassl2021}.

The lower bound in common use is the Brun--Devetak--Hsieh (BDH) form of
the entanglement-assisted Singleton bound, $2(d-1)\le n-k+c$~\cite{BDH2014}.
It holds for every code with $d-1\le n/2$~\cite{LaiAshikhmin2018,GHW2022}.
Distances with $d-1\ge n/2$, at which half of the qudits or more can be
lost, require entanglement assistance. Without it, two disjoint halves of
the qudits would each determine the logical qudits, which the no-cloning
theorem forbids. For $d-1>n/2$, Grassl~\cite{Grassl2021} and our earlier
work~\cite{BaekWonKim2026} constructed codes that violate the BDH bound.
Grassl, Huber and Winter proved a generalized bound~\cite{GHW2022} that is
much weaker at large distance. Two questions were open: \emph{For
which parameters is the BDH bound valid? By how much can it overestimate
the entanglement?}

We answer both questions for $\ea{n}{k}{d}{c}_q$ codes. Our approach
starts from erasures, losses of qudits at known positions. A code of
distance $d$ corrects $d-1$ erasures, so any $n-d+1$ of its qudits, together with the receiver's
halves of the pairs, determine the $k$ logical qudits. The answer depends
on how this number compares with $2k$, that is, on how $k$ compares with
the Singleton defect $\sigma=n-k-d+1$. When $k>\sigma$, a count of
dimensions on such sets of qudits proves the BDH bound, for degenerate
codes as well. When $k=\sigma$ the bound can fail, and the hull of a code
that violates it contains a classical MDS code of dimension $2k$. The
violation is therefore limited by the length of the longest such code,
and codes built on the normal rational curve and its tangent lines reach
this limit. When $k<\sigma$ the same argument leads to classical codes of
larger Singleton defect.

Section~\ref{sec:prelim} sets up the codes, Sec.~\ref{sec:main} treats the
three cases, and Sec.~\ref{sec:budget} discusses what the bounds imply for
entanglement cost and compares them with the tables of best-known codes.
Appendix~\ref{app:notation} fixes the notation. Appendix~\ref{app:proofs}
contains the proofs and shows how the results depend on each other. All of
them, except for cited lengths of classical MDS codes, are formalized in the
Lean~4 proof assistant. Appendix~\ref{app:examples} gives examples and
comparisons, and Appendix~\ref{sec:machine} describes the machine checks
and the formalization.

\section{Entanglement-assisted stabilizer codes}\label{sec:prelim}

A qudit Pauli operator on $n$ qudits of prime-power dimension $q$
(modulo phase) is $v\in\F_q^{2n}$, acting on qudit $i$ as
$X^{x_i}Z^{z_i}$. The symplectic form is
$\langle u,v\rangle=\sum_i (x_iz_i'-z_ix_i')\in\F_q$, and $\wt(v)$
counts the qudits $i$ at which the component $(x_i,z_i)$ of $v$ is
nonzero. For an $\F_q$-linear code $\Cc\le\F_q^{2n}$ we write $\Cd$ for
its symplectic dual and $\Hull=\Cc\cap\Cd$ for its hull, following Galindo
\emph{et al.}~\cite{Galindo2019}. If $\dim\Cc=n-k+c$ and
$\dim\Hull=n-k-c$, then $\Cc$ is the stabilizer of an EA stabilizer code
with $k$ logical qudits, $c$ pairs and distance
$d=\min\{\wt(v)\mid v\in\Cd\setminus\Hull\}$~\cite{WildeBrun2008,Galindo2019,Luo2022},
and $\dim\Cd-\dim\Hull=2k$. For a prime power $q=p^h$, with $p$ prime and $h\ge1$, the physical
commutation relation is the trace-symplectic one,
$\mathrm{Tr}_{\F_q/\F_p}\langle u,v\rangle=0$. For $\F_q$-linear codes the
two conditions agree, because $\mathrm{Tr}(\lambda\langle u,v\rangle)=0$
for every $\lambda\in\F_q$ forces $\langle u,v\rangle=0$~\cite{Ketkar2006}.

In the language of Pauli operators, $\Cc$ is the stabilizer group
restricted to the sender's $n$ qudits, and its elements need not commute.
The elements of $\Hull$ commute with all of $\Cc$ and need no
entanglement. The other generators of $\Cc$ can be chosen as $c$ pairs of
operators that do not commute, and each such pair is made to commute by one
entangled pair shared with the receiver~\cite{BDH2006,WildeBrun2008}. The
elements of $\Cd$ are the Pauli operators that commute with all of $\Cc$.
Those outside $\Hull$ act nontrivially on the logical qudits, so $d$ is the
least number of qudits on which a nontrivial logical operator acts. A code
is degenerate if a nonzero element of $\Hull$ acts on fewer than $d$ qudits.

We write $\ea{n}{k}{d}{c}_q$ for such a code. The brackets indicate that
$\Cc$ is $\F_q$-linear. For prime $q$ every additive code is
$\F_q$-linear. For other $q$ an additive code that is not $\F_q$-linear is
written $\eaa{n}{q^k}{d}{c}_q$. Every code below except the additive
example of Sec.~\ref{sec:bdh} is an $\ea{n}{k}{d}{c}_q$ code. So is every
code that our computer searches range over, because they enumerate
$\F_q$-linear subspaces, not arbitrary encoders. Following Luo, Ezerman, Grassl and Ling~\cite{Luo2022}, we
define the least entanglement of such codes as
\begin{equation}\label{eq:cmin}
  \cmins(n,k,d)=\min\bigl\{c\bigm|\text{an }\ea{n}{k}{d}{c}_q\text{ code exists}\bigr\},
\end{equation}
where the distance $d$ is itself a minimum, the least weight of a vector
of $\Cd\setminus\Hull$. All lower bounds below are lower bounds on
$\cmins(n,k,d)$. Entanglement-assisted codes with a nonadditive encoder
fall outside the formalism. They exist and have been constructed
explicitly~\cite{Shin2011}, so $\cmins$ is in general only an upper bound
for the minimum over all EAQECCs.

\begin{table}[htbp]
\caption{Notation.}
\label{tab:notation}
\setlength{\tabcolsep}{3pt}
\centering
\begin{tabular}{ll}
\toprule
$\sigma=n-k-d+1$ & Singleton defect\\
$\Delta=2(d-1)-(n-k+c)$ & \begin{tabular}[t]{@{}l@{}}violation of the BDH bound,\\which reads $\Delta\le0$\end{tabular}\\
$\cmins(n,k,d)$ & \begin{tabular}[t]{@{}l@{}}least $c$ of an $\ea{n}{k}{d}{c}_q$ code,\\Eq.~(\ref{eq:cmin})\end{tabular}\\
\bottomrule
\end{tabular}
\end{table}

We write $\sigma=n-k-d+1$ for the Singleton defect (Table~\ref{tab:notation}).
The number of qudits that, with the receiver's halves of the pairs,
determine the logical qudits is $n-d+1=k+\sigma$. The cases $k>\sigma$,
$k=\sigma$ and $k<\sigma$ are those in which it is less than, equal to or
greater than $2k$. The \emph{violation} of the BDH bound
$2(d-1)\le n-k+c$ is
\begin{equation}\label{eq:excess}
  \Delta=2(d-1)-(n-k+c)=\dim\Hull-2\sigma .
\end{equation}
It is the number of pairs a code uses
fewer than that bound requires, and the bound reads $\Delta\le0$. The
second expression follows from $\dim\Hull=n-k-c$ and
$2\sigma=2(n-k)-2(d-1)$:
\begin{equation*}
\begin{aligned}
  \dim\Hull-2\sigma&=(n-k-c)-2(n-k)+2(d-1)\\
  &=2(d-1)-(n-k+c).
\end{aligned}
\end{equation*}
For
$d-1<n/2$ the codes with $\Delta=0$ are the entanglement-assisted quantum
MDS (EAQMDS) codes of Grassl, Huber and Winter~\cite{GHW2022}. For
$d-1\ge n/2$ and $k\ge1$ their generalized Singleton
bound~\cite{GHW2022} reads
\begin{equation}\label{eq:ghw}
\begin{gathered}
k \le \frac{n-d+1}{3(d-1)-n}\bigl(c+2(d-1)-n\bigr),\\
k\le c,\qquad k\le n-d+1 ,
\end{gathered}
\end{equation}
and every code constructed below satisfies it.

\section{Validity and violation of the BDH bound}\label{sec:main}

We treat the cases $k>\sigma$, $k=\sigma$ and $k<\sigma$, in which fewer
than $2k$, exactly $2k$ or more than $2k$ qudits, with the receiver's
halves of the pairs, determine the logical qudits.
Table~\ref{tab:summary} collects the results.

\begin{table}[htbp]
\centering
\setlength{\tabcolsep}{1.8pt}
\begin{lrbox}{\tablebody}%
\begin{tabular}{llll}
\toprule
Case & $k$, $q$ & Largest $\Delta$ & By\\
\hline
$k>\sigma$ & all & $\le0$ & Thm.~\ref{thm:bdh}\\
$k=\sigma$ & $k=1$ & $q-1$ & Cor.~\ref{cor:diagexact}\\
 & $k=2$, $q\ge4$ & $q-3$ & Cor.~\ref{cor:diagexact}\\
 & $k=2$, $q=2$ & $\le0$ & Prop.~\ref{prop:k2}\\
 & $k=2$, $q=3$ & $0$ & Prop.~\ref{prop:k2}\\
 & $2k{\le}q$, $M_q(2k){=}q{+}1^{\dagger}$ & $q+1-2k$ & Cor.~\ref{cor:diagexact}\\
 & other $2k\le q$ & $\le M_q(2k)-2k^{\ddagger}$ & Thm.~\ref{thm:diag}\\
 & $2k>q$ & $\le1$ & Cor.~\ref{cor:diag}\\
$k<\sigma$ & $k=1$ & $(q-1)\sigma$ & Cor.~\ref{cor:k1}\\
 & $k\ge2$ & $\le B_q(k,\sigma)-2\sigma$ & Thm.~\ref{thm:above}\\
\bottomrule
\end{tabular}%
\end{lrbox}
\begin{minipage}{\wd\tablebody}
\centering
\caption{The largest violation $\Delta$ by case.}
\label{tab:summary}
\noindent\usebox{\tablebody}\par\smallskip
\footnotesize\raggedright
$^{\dagger}$Including every prime $q$, every $q\le31$ and $q=49$, and every even $q$ by the results announced by Chen and Huang~\cite{ChenHuang2026} and by Kumar~\cite{Kumar2026}. Appendix~\ref{app:mq} lists all known cases.\par
$^{\ddagger}$At least $q+1-2k$ by Theorem~\ref{thm:s-full-range}, and equal to it under the MDS conjecture.\par
$M_q(2k)$ is defined in Sec.~\ref{sec:diag} and $B_q(k,\sigma)$ in Sec.~\ref{sec:further}.\par
\end{minipage}
\end{table}

\subsection{The case \texorpdfstring{$k>\sigma$}{k > σ}}\label{sec:bdh}

The BDH bound was known to hold for every code with
$d-1\le n/2$~\cite[Cor.~9]{GHW2022}, and at every $d$ for nondegenerate
codes~\cite{LaiAshikhmin2018,Nadkarni2021}, pure codes~\cite{GHW2022},
$\F_{q^2}$-linear codes, where it is the relative Singleton
bound~\cite{LuoRGHW2005}, and the Hermitian and CSS-like
constructions~\cite{Luo2022}. Lai and Ashikhmin~\cite{LaiAshikhmin2018} left
the general stabilizer case $d-1>n/2$ open and proved there a refined
Singleton bound of linear-programming type for general qubit codes.
Appendix~\ref{app:bdh-ex} compares it with our bounds at the parameters they
evaluate. The difficulty lies in
degenerate codes, whose hull may contain vectors of weight less than $d$.
When $k>\sigma$, a count of dimensions on sets of $n-d+1$ qudits removes it.

\begin{theorem}[Validity for $k>\sigma$]\label{thm:bdh}
Every $\ea{n}{k}{d}{c}_q$ code with $k>\sigma$ satisfies $\Delta\le0$, the BDH
bound.
\end{theorem}

The proof (Appendix~\ref{app:bdh}) restricts Pauli operators to a set $T$
of $k+\sigma=n-d+1$ qudits. An element of $\Cd$ that acts trivially on $T$
acts on at most $d-1$ qudits, so it lies in $\Hull$ and acts trivially on
the logical qudits. Logical operators that act differently on the logical
qudits therefore differ on $T$. Their restrictions to $T$ span $2k$
dimensions beyond the restriction of $\Hull$, out of the $2(k+\sigma)$
dimensions of Pauli operators on $T$, so the restriction of $\Hull$ to $T$
has dimension at most $2\sigma$ (Lemma~\ref{lem:perset}). Only the weights
of logical operators enter, not those of elements of $\Hull$, so the count
holds for degenerate codes. Choosing qudits one at a time, each raising the
dimension of the restriction of $\Hull$, reaches dimension
$\min(\dim\Hull,2\sigma+1)$ with at most $2\sigma+1$ qudits. When $k>\sigma$
these fit inside a set of $k+\sigma$ qudits, so $\dim\Hull\le2\sigma$,
which is $\Delta\le0$.

Theorem~\ref{thm:bdh} proves the BDH bound for $k>\sigma$ also when
$d-1>n/2$, the case left open by Lai and Ashikhmin~\cite{LaiAshikhmin2018}. For prime $q$,
qubits in particular, every EA stabilizer code is an $\ea{n}{k}{d}{c}_q$
code (Sec.~\ref{sec:prelim}), so there the theorem covers all EA stabilizer
codes with $k>\sigma$. Neither $k>\sigma$ nor $\F_q$-linearity
can be dropped. For $k=\sigma$
the codes of Sec.~\ref{sec:diagcodes} and Grassl's teleportation
codes~\cite{Grassl2021} violate the bound, and an additive
$\eaa{10}{121^3}{7}{4}_{121}$ code that is $\F_{11}$-linear but not
$\F_{121}$-linear, with $k=3>\sigma=1$, has $\Delta=1$
(Appendix~\ref{app:bdh-ex}).

\subsection{The case \texorpdfstring{$k=\sigma$}{k = σ}}\label{sec:diag}

\begin{figure}[htbp]
\centering
\resizebox{\linewidth}{!}{\input{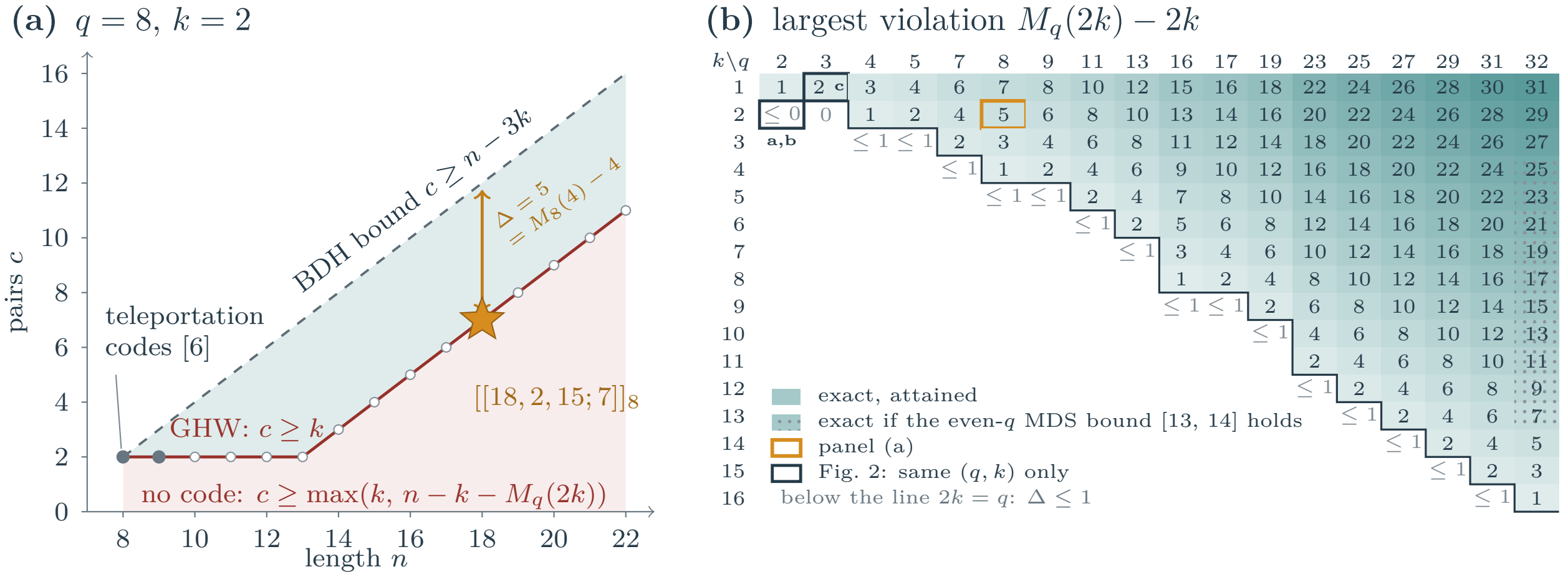}}%
\caption{Violation of the BDH bound for $k=\sigma$. (a)~Theorem~\ref{thm:diag}
and the bound of Grassl, Huber and Winter (GHW)~\cite{GHW2022} exclude the
red region. The star is the code of Theorem~\ref{thm:diagcodes}. Open
circles mark open lengths.}
\label{fig:cost}
\end{figure}

At $k=\sigma$ exactly $2k$ qudits, with the receiver's halves of the pairs,
determine the logical qudits, and the BDH bound can fail. Here it demands
$c\ge n-3k$, while the generalized bound of Grassl, Huber and
Winter~\cite{GHW2022} gives only $c\ge k$. For $q=8$, $k=2$, $n=18$ and
$d=15$ the two bounds demand $12$ pairs and $2$ pairs. We bound the
violation by the length of a classical MDS code of dimension $2k$ and
construct codes that attain this bound. A classical $[m,\ell]_q$ code, of
length $m$ and dimension $\ell$, is maximum distance separable (MDS) if any
$\ell$ of its $m$ symbols determine the encoded message. We use $m$ and
$\ell$ rather than $n$ and $k$ because the classical codes below have
length $\dim\Hull$ and dimension $2k$. Extended Reed--Solomon codes, which evaluate
polynomials of degree less than $\ell$ at the $q+1$ points of the projective
line over $\F_q$, are MDS codes of length $q+1$. For $\ell\ge2$ let
$M_q(\ell)$ be the largest $m$ for which an MDS $[m,\ell]_q$ code exists.

\subsubsection{Upper bound on the violation}

\begin{theorem}[Upper bound on the violation for $k=\sigma$]\label{thm:diag}
Every $\ea{n}{k}{d}{c}_q$ code with $k=\sigma\ge1$ has
$\dim\Hull\le M_q(2k)$, so $\Delta\le M_q(2k)-2k$.
\end{theorem}

The proof (Appendix~\ref{app:diag}) has two steps. If at some qudit the
components of the vectors of $\Hull$ span $\F_q^2$, intersecting $\Cd$ and
$\Hull$ with the kernel of that qudit's projection and deleting the qudit
gives a code with the same $k$, distance at least $d$, a hull smaller by
two dimensions and Singleton defect at most $\sigma-1$, to which
Theorem~\ref{thm:bdh} applies. Otherwise each qudit in
the support of $\Hull$ defines a linear functional on
$\Cd/\Hull\cong\F_q^{2k}$, and any $2k$ of these functionals at qudits
with independent hull coordinates are linearly independent. The cosets of
$\Hull$ in $\Cd$ are the logical classes. In words,
each such qudit carries one linear function of the logical class, as a
symbol of a classical code carries one linear function of its message. The
functionals are then the columns of an MDS code of dimension $2k$ and
length $\dim\Hull$. Equivalently, they form an arc of
$\mathrm{PG}(2k-1,q)$.

The proof is constructive. For an $\ea{n}{k}{n-2k+1}{c}_q$ code with
$\Delta>0$ the first step does not occur, and the functionals, read off
from generators of $\Cd$, give a linear $[n-k-c,\,2k]_q$ MDS code.
Lowering $c$ at fixed $n$, $k$ and $d=n-2k+1$ therefore requires a longer
MDS code of dimension $2k$. At $n=2q+2$, a code with $c<q+1-k$ would give
an MDS code of length $q+2$ and dimension $2k\le q$, contrary to the MDS
conjecture and, in the cases listed below, impossible.

The MDS conjecture predicts $M_q(\ell)=q+1$ for all even $\ell\le q$,
while $M_q(\ell)=\ell+1$ for $\ell\ge q$. Collecting the known cases
(Appendix~\ref{app:mq}), $M_q(2k)=q+1$ for every $2k\le q$ when $q$ is
prime~\cite{Ball2012}, when $q\le31$~\cite{BallLavrauw2019,HS2001} and
when $q=49$~\cite{Fan2026}. Chen and Huang~\cite{ChenHuang2026} and,
independently, Kumar~\cite{Kumar2026} have announced proofs of the MDS
conjecture for every even $q$, which give $M_q(2k)=q+1$ for every even $q$
and $2k\le q$. With them the first open case is $q=81$ with
$30\le2k\le52$. In
these cases Theorem~\ref{thm:diag} gives $\Delta\le q+1-2k$, and for
$2k\ge q$ it gives $\Delta\le1$ (Corollary~\ref{cor:diag}).

\subsubsection{Codes attaining the bound}\label{sec:diagcodes}

For $2k\le q$ we now construct codes with $\Delta=q+1-2k$. Take one pair
of qudits for each point $s$ of $\mathbb{P}^1(\F_q)=\F_q\cup\{\infty\}$, so
$n=2q+2$, and let $\rho_s$ be the vector with $z$-components $(1,-1)$ on
the two qudits of pair $s$ and zeros elsewhere, the operator
$Z\otimes Z^{-1}$ on that pair. Define logical generators
$L_0,\dots,L_{2k-1}$ by
\begin{equation}\label{eq:diagcode}
\begin{aligned}
s\in\F_q:&\quad x=(s^r,s^r),\quad z=(\eta_s\,r\,s^{r-1},0),\\
\infty:&\quad x=(\chi_r,\chi_r),\quad z=(\delta\,\chi'_r,0)
\end{aligned}
\end{equation}
on the two qudits of pair $s$. Here $\chi_r$ is $1$ if $r=2k-1$ and $0$
otherwise, and $\chi'_r$ is $1$ if $r=2k-2$ and $0$ otherwise. We read
$0^0$ as $1$ and $rs^{r-1}$ as $0$ at $r=0$. The scalars are
$\eta_s=s^{-(2k-2)}$ for $s\ne0$ and $\eta_0=1$, and
$\delta\in\F_q^{\times}$ is chosen below. Let $\Cc$ be the symplectic dual
of $\langle L_0,\dots,L_{2k-1},\rho_s\mid s\in\mathbb{P}^1(\F_q)\rangle$, so
that $\Cd$ is this span. A logical class contains exactly one vector $\sum_ra_rL_r$, and we identify it with
$a\in\F_q^{2k}$ and with the polynomial $p_a(x)=\sum_{r<2k}a_rx^r$.

\begin{theorem}[Codes attaining the bound]\label{thm:diagcodes}
Let $q=p^h$ and $1\le k\le q/2$, and suppose $p=2$ or $p$ divides no
positive odd integer $m\le2k-5$ (automatic for prime $q$). Then for a
suitable $\delta$ the code $\Cc$ of~\eqref{eq:diagcode} is a
$\ea{2q+2}{k}{2q+3-2k}{q+1-k}_q$ code, with $\Delta=q+1-2k$.
\end{theorem}

The proof is in Appendix~\ref{app:codes}. Its distance step uses that a
nonzero polynomial of degree less than $2k$ has at most $2k-1$ roots on
$\mathbb{P}^1(\F_q)$, counted with multiplicity, and that a vector of least
weight in its coset has one zero component on the pair of a simple root
and two on the pair of a multiple root. Geometrically, the functionals in
the proof of Theorem~\ref{thm:diag} are here the points
$(1,s,\dots,s^{2k-1})$ of the normal rational curve of
$\mathrm{PG}(2k-1,q)$, each carried by both qudits of pair $s$, and a
logical operator, viewed as a hyperplane, vanishes on both qudits of a pair
exactly when it contains the tangent line of the curve there. The arc that
Theorem~\ref{thm:diag} extracts from any code is here the normal rational
curve, of the largest size $q+1$ whenever $M_q(2k)=q+1$. With suitable
nonzero scalars in place of the closed formula, the same generators give
these parameters for every $q$ and $1\le k\le q+1$
(Theorem~\ref{thm:s-full-range} in Appendix~\ref{app:codes}).

\begin{corollary}[Largest violation for $k=\sigma$]\label{cor:diagexact}
Let $q=p^h\ge2k$ and $k\ge1$. If $M_q(2k)=q+1$, as in the cases listed in
Appendix~\ref{app:mq}, then the largest violation of an
$\ea{n}{k}{d}{c}_q$ code with $k=\sigma$ is exactly $q+1-2k$, attained at
$n=2q+2$.
\end{corollary}

For even $q$ the condition of
Theorem~\ref{thm:diagcodes} always holds, so the codes~\eqref{eq:diagcode}
exist for every even $q$ and $2k\le q$, and we checked them explicitly for
$q=32$ and $q=64$ and every $k\le q/2$
(Appendix~\ref{sec:machine}). With the results announced for even $q$,
Corollary~\ref{cor:diagexact} therefore gives
the largest violation for $k=\sigma$ and $2k\le q$ for every prime $q$,
every even $q$ and every $q<81$. In the example above,
Theorem~\ref{thm:diag} gives $c\ge7$, and the code $\ea{18}{2}{15}{7}_8$ of
Theorem~\ref{thm:diagcodes} attains it. The BDH bound overstates the least
entanglement and the generalized bound understates it, here by five pairs
each.

The smallest case, $q=2$ and $k=1$, is a code on six qubits. Pairs $0$, $1$
and $\infty$ are qubits $1,2$, qubits $3,4$ and qubits $5,6$, and the hull is
generated by $Z_1Z_2$, $Z_3Z_4$ and $Z_5Z_6$. The logical generators of
Eq.~\eqref{eq:diagcode} are $L_0=X_1X_2X_3X_4Z_5$ and $L_1=Z_1Y_3X_4X_5X_6$,
which do not commute. The stabilizer $\Cc$, the set of Pauli operators that
commute with these five, has dimension $7$, so $c=2$. Both qubits of pair
$s$ carry the $X$-exponent $p_a(s)=a_0+a_1s$ of the logical operator
$a_0L_0+a_1L_1$, with $a_1$ at $\infty$. A nonzero $p_a$ vanishes at one
point of $\mathbb{P}^1(\F_2)$ at most. On that pair the logical operator
acts on one qubit, and multiplying it by the hull generator of the pair
cannot clear both. Every logical operator therefore acts on at least five
qubits, and $d=5$. Any two qubits, with the receiver's two qubits, thus
determine the logical qubit, which no code without entanglement allows. The
BDH bound demands $c\ge3$ and the bound of Grassl, Huber and Winter only
$c\ge1$. In the proof of Theorem~\ref{thm:diag} the functional of a qubit of
pair $s$ is $a\mapsto p_a(s)$, so the three pairs give the columns $(1,0)$,
$(1,1)$ and $(0,1)$ of the binary $[3,2,2]$ even-weight code. It is the
longest binary MDS code of dimension two, and Theorem~\ref{thm:diag} gives
$c\ge n-k-M_2(2)=2$. The code attains this and violates the BDH bound by
one pair.

\begin{table}[htbp]
\centering
\setlength{\tabcolsep}{2.5pt}
\begin{lrbox}{\tablebody}%
\begin{tabular}{lll}
\toprule
Parameters & $k$ & Distance established by\\
\hline
$\ea{6}{2}{3}{1}_{2}$, $\ea{7}{2}{4}{2}_{2}$ & $2$ & exhaustive$^{a}$\\
$\ea{9}{2}{5}{2}_{2}$ & $2$ & exhaustive\\
$\ea{10}{2}{6}{3}_{3}$, $\ea{12}{2}{8}{5}_{3}$ & $2$ & exhaustive\\
$\ea{10}{k}{11-2k}{5-k}_{4}$ & $2,3$ & exhaustive, G at $3$\\
$\ea{12}{k}{13-2k}{6-k}_{5}$ & $2,3$ & exhaustive, G at $3$\\
$\ea{16}{k}{17-2k}{8-k}_{7}$ & $2,\dots,4$ & class by class, G at $4$\\
$\ea{18}{k}{19-2k}{9-k}_{8}$ & $2,\dots,5$ & class by class, G at $5$\\
$\ea{20}{k}{21-2k}{10-k}_{9}$ & $2,3,4^{b},5$ & class by class, G at $5$\\
$\ea{24}{k}{25-2k}{12-k}_{11}$ & $2,\dots,6$ & class by class, G at $6$\\
$\ea{28}{k}{29-2k}{14-k}_{13}$ & $2,\dots,7$ & class by class, G at $6,7$\\
\bottomrule
\end{tabular}%
\end{lrbox}
\begin{minipage}{\wd\tablebody}
\centering
\caption{Concrete codes with $k\ge2$, by field size.}
\label{tab:codes}
\noindent\usebox{\tablebody}\par\smallskip
\footnotesize\raggedright
Exhaustive: least weight over all of $\Cd\setminus\Hull$. Class by class: least weight in each coset of $\Hull$. G: exact Gram rank, a vector of least weight and the root-counting lower bound.\par
$^{a}$Found with a SAT solver. The first is in~\cite{LaiBrun2013}, and both are among the records released with~\cite{BaekWonKim2026}. $^{b}$With scalars found by computer search (Table~\ref{tab:s-scalars}).\par
\end{minipage}
\end{table}

\begin{table}[htbp]
\centering
\setlength{\tabcolsep}{3pt}
\begin{lrbox}{\tablebody}%
\begin{tabular}{lll}
\toprule
$q$ & $k$ & Established by\\
\hline
All & $1\le k\le q+1$ & Theorem~\ref{thm:s-full-range}\\
$q\le13$ & $1\le k\le q+1$ & G, all $71$ cases $(q,k)^{a}$\\
$16,25,27$ & $2$ & class by class\\
$16$ & $4$ & class by class\\
$27$ & $4^{b}$ & class by class\\
$32$, $64$ & $1\le k\le q/2$ & G\\
\bottomrule
\end{tabular}%
\end{lrbox}
\captionsetup{width=.84\linewidth,justification=centering}
\caption{The codes \mbox{$\ea{2q+2}{k}{2q+3-2k}{q+1-k}_q$} of Eq.~\eqref{eq:diagcode} with suitable nonzero scalars.}
\label{tab:fullrange}
\begin{minipage}{\wd\tablebody}
\centering
\noindent\usebox{\tablebody}\par\smallskip
\footnotesize\raggedright
G, class by class: as in Table~\ref{tab:codes}. $^{a}$Of these, $24$ small cases are also verified class by class. $^{b}$With scalars found by computer search (Table~\ref{tab:s-scalars}). The other rows with $q>13$ use Theorem~\ref{thm:diagcodes}.\par
\end{minipage}
\end{table}

\begin{figure}[htbp]
\centering
\resizebox{\linewidth}{!}{\input{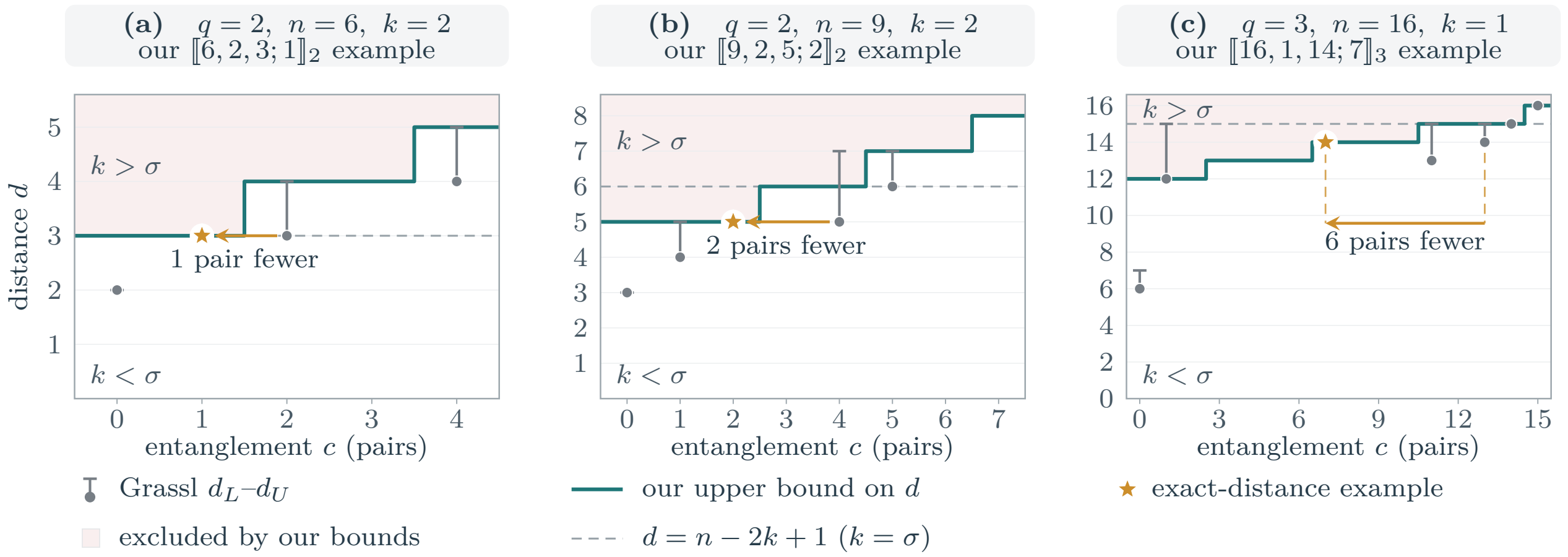}}%
\caption{Entanglement--distance tradeoffs at fixed $(q,n,k)$. Grassl's lower
and upper distances $d_L$ and $d_U$~\cite{codetables} include
teleportation-based codes. Our upper bounds come from
Theorem~\ref{thm:bdh}, Proposition~\ref{prop:k2} and
Corollary~\ref{cor:k1}. Arrows compare with the fewest pairs listed for the
same distance.}
\label{fig:map}
\end{figure}

Figure~\ref{fig:cost}(a) shows the least entanglement for $k=\sigma$ as a
function of the length. At other lengths the bound of Grassl, Huber and
Winter~\cite[Cor.~9]{GHW2022} gives $\Delta\le\max(0,n-4k)$ for
$k=\sigma$, and Grassl's teleportation codes from MDS
codes~\cite{Grassl2021}, which are $\ea{n}{k}{d}{c}_q$ codes, attain
$n-4k$ for $4k\le n\le M_q(2k)$. With Theorem~\ref{thm:diag} the least
entanglement for $k=\sigma$ is therefore at least
$\max(k,\,n-k-M_q(2k))$ for every $n\ge4k$, attained by teleportation codes
at short lengths and by our codes at $n=2q+2$. For $k\ge2$ the lengths
$M_q(2k)<n<2q+2$ and $n>2q+2$ are open, and whether $\Delta=1$ can occur
when $2k>q$ remains open in general. Figure~\ref{fig:cost}(b) shows the
largest violation for $q\le32$.

Tables~\ref{tab:codes} and~\ref{tab:fullrange} list concrete codes and
how their parameters were established.
Table~\ref{tab:fullrange} gives the codes~\eqref{eq:diagcode} with suitable
nonzero scalars for every $1\le k\le q+1$, followed by larger fields. For
the codes~\eqref{eq:diagcode} with $\Delta\ge-1$, the listed $d$ is the
largest possible by Theorem~\ref{thm:bdh}, since one more unit of distance
makes $\sigma<k$ and the violation positive. For the other codes with
$k=2$ over $\F_2$ and $\F_3$, one more unit of distance gives $\sigma=1<k$
or $k=\sigma=2$ with $\Delta=1$, which Theorem~\ref{thm:bdh} or
Proposition~\ref{prop:k2} excludes. For the remaining codes the listed $d$
is the exact distance of the listed code, and a larger distance with the
same $n$, $k$ and $c$ is not excluded. Whenever $2k\le q$ in
Tables~\ref{tab:codes} and~\ref{tab:fullrange}, Corollary~\ref{cor:diagexact} shows in addition that
the listed $c$ is the least possible for the listed $n$, $k$ and $d$. In
Table~\ref{tab:fullrange}, Theorem~\ref{thm:s-full-range} proves the first
row for every finite field, and the remaining rows record computations that
verify them (Appendix~\ref{sec:machine}).

\subsubsection{One logical qudit and small fields}\label{sec:small}

The bounds above use only dimensions. That $\Hull$ is orthogonal to $\Cd$
gives in addition an identity between the functionals in the proof of
Theorem~\ref{thm:diag} (Appendix~\ref{sec:length}), with two consequences.
For one logical qudit with $k=\sigma$, the least entanglement
$\cmins(n,1,n-1)$ is determined at every length $n\ge2q+1$, by
Theorem~\ref{thm:k1length} and the codes of
Propositions~\ref{thm:floorfam} and~\ref{thm:unified}. The codes of
Proposition~\ref{thm:unified} are those of Theorem~1 of our earlier
work~\cite{BaekWonKim2026}, whose Sec.~4.3 obtains these values by computer
for $q=2$ with $6\le n\le11$ and for $q=3$ with $n=10$.

For $k=\sigma=2$ and $q\le3$ no code violates the BDH bound
(Proposition~\ref{prop:k2}). The proof is computer assisted. The identity
and a counting argument reduce the question to finitely many Boolean
satisfiability (SAT) problems, one for each possible size of the support
of the hull. The SAT solver CaDiCaL~\cite{CaDiCaL} finds each of them unsatisfiable and
writes a proof of this in the DRAT format, which the independent checker
\texttt{drat-trim}~\cite{DRATtrim} verifies (Appendix~\ref{sec:length}). The reduction,
the encodings and the refutations are also checked by the Lean kernel
(Appendix~\ref{sec:machine}). For odd $q$ the codes~\eqref{eq:diagcode}
with $k=(q+1)/2$ and suitable scalars (Theorem~\ref{thm:s-full-range}) give
$\Delta=0$. At $q=3$ this is the code $\ea{8}{2}{5}{2}_3$, which attains
the bound of Proposition~\ref{prop:k2}.

\subsection{The case \texorpdfstring{$k<\sigma$}{k < σ}}\label{sec:further}

When $k<\sigma$, more than $2k$ qudits are needed to determine the logical
qudits, and the bound involves classical codes of positive Singleton
defect. The Singleton defect of an $[m,\ell]_q$ code is the least
$\epsilon$ such that any $\ell+\epsilon$ of its $m$ symbols determine the
encoded message. Let $N_q(\ell,\epsilon)$ be the largest length of a $q$-ary
linear code of dimension $\ell\ge2$ and Singleton defect at most
$\epsilon$. Define
\begin{equation*}
  B_q(k,\sigma)=\begin{cases}
  2\sigma & \text{if } \sigma<k,\\
  \max\{N_q(2k,\sigma-k),\,2+B_q(k,\sigma-1)\} & \text{if } \sigma\ge k.
  \end{cases}
\end{equation*}

\begin{theorem}[Bound for $k<\sigma$]\label{thm:above}
Every $\ea{n}{k}{d}{c}_q$ code has $\dim\Hull\le B_q(k,\sigma)$, so
$\Delta\le B_q(k,\sigma)-2\sigma$. If moreover $\sigma\ge k$ and at no
qudit the components of the vectors of $\Hull$ span $\F_q^2$, then
$\dim\Hull\le N_q(2k,\sigma-k)$.
\end{theorem}

The proof (Appendix~\ref{sec:above}) repeats the two steps of
Theorem~\ref{thm:diag}. Each shortening at a qudit lowers $\sigma$ by at
least one, and when no qudit is left to shorten, the functionals form a
classical code of dimension $2k$ and Singleton defect at most
$\sigma-k$. For $k=1$ the
Griesmer bound gives $B_q(1,\sigma)=(q+1)\sigma$, so
$\Delta\le(q-1)\sigma$, which is attained (Corollary~\ref{cor:k1}). The
violation can thus grow linearly with $\sigma$.

\section{Implications for entanglement cost}\label{sec:budget}

Read as statements about resources, the results bound the pairs that a
linear EA stabilizer scheme must consume, however it is built.
Teleportation codes~\cite{Grassl2021} are of this kind. Whenever
$k>\sigma$, the BDH bound $c\ge2(d-1)-(n-k)$ is valid
(Theorem~\ref{thm:bdh}). For $k=\sigma$ a code of length $n$ violates it
by at most $\min(\max(0,n-4k),\,M_q(2k)-2k)$ pairs, which
is at most $q+1-2k$ whenever $M_q(2k)=q+1$ (Fig.~\ref{fig:cost}). For
$k<\sigma$ the violation is at most $B_q(k,\sigma)-2\sigma$
(Theorem~\ref{thm:above}), which for $k=1$ is exactly $(q-1)\sigma$.

The natural comparison is with teleportation. The sender teleports the
$k$ qudits with $k$ pairs and protects the $2k$ classical outcomes with a
classical $[n,2k,d]_q$ code. Grassl's teleportation codes realize this as
$\ea{n}{k}{d}{k}_q$ codes~\cite{Grassl2021}. For $k=\sigma$ and
$4k\le n\le M_q(2k)$, built from an MDS code, they use the fewest pairs
possible, since the bound of Grassl, Huber and Winter gives $c\ge k$. For
longer codes Theorem~\ref{thm:diag} shows that every linear scheme,
teleportation-based or not, needs $c\ge n-k-M_q(2k)$. At $n=2q+2$ and
$d=2q+3-2k$ this is $q+1-k$ whenever $M_q(2k)=q+1$. This distance at this
length therefore costs $q+1-2k$ pairs beyond teleportation, and the codes of
Theorem~\ref{thm:diagcodes} use exactly that many. They show that the bound
of Theorem~\ref{thm:diag} is attained, not that they are the cheapest way to
send $k$ qudits, and their cost beyond teleportation is small only when $2k$
is close to $q$.

The same parameters describe the storage of $k$ qudits in $n$ qudit
memories such that any $n-d+1$ of them, together with $c$ reference qudits
kept apart, recover the stored state. The capacity of such storage when
the assistance can also be erased was characterized by Sun and
Jafar~\cite{SunJafar2026}, apart from an intermediate range. The results
here concern a fixed alphabet, finite length and a reliable reference.
Figure~\ref{fig:map} places our upper bounds on $d$ next to Grassl's tables of
best-known qubit and qutrit codes, and Appendix~\ref{sec:machine} checks
every listed code against them. Compared with these tables as of 15 October
2026, and with the published codes they omit~\cite{LaiBrun2013,LiRenGuanLiu2026,BaekWonKim2026},
the codes of this paper use fewer pairs at the same distance in $61$ cases, by up to $20$
pairs, and raise the best known distance in $74$ listed entries
(Appendix~\ref{sec:machine}).

\section{Concluding remarks}

For $\ea{n}{k}{d}{c}_q$ codes the BDH bound is valid whenever $k>\sigma$,
degenerate codes included. For $k=\sigma$ the violation is at most
$M_q(2k)-2k$ pairs. When $2k\le q$ and $M_q(2k)=q+1$, as the MDS conjecture
predicts for every $q$, the largest violation is $q+1-2k$. The
entanglement these codes need is thus tied to the lengths of classical MDS
codes. All bounds come from the
dimension count of Lemma~\ref{lem:perset}, applied to a greedily chosen set
of qudits, to a single qudit at which the hull projects onto $\F_q^2$, and
to the $k+\sigma$ qudits of an arc. Only the results for one logical qudit
and small fields also use that $\Hull$ is orthogonal to $\Cd$.

Several questions remain open. For $k=\sigma$ and $k\ge2$ the least
entanglement is unknown between the two ends of Fig.~\ref{fig:cost}(a) and
beyond $n=2q+2$. For $2k>q$ only $\Delta\le1$ is known, apart from
$\Delta\le0$ for $q\le3$ and $k=2$. The code $\ea{8}{2}{5}{2}_3$ attains
this bound, and binary attainment is open. For additive codes that are not
$\F_q$-linear, over composite fields, Theorem~\ref{thm:bdh} fails and the
right condition is unknown. It is also open whether $B_q(k,\sigma)$ is
attained for $2\le k<\sigma$, and whether nonadditive
codes~\cite{Shin2011,DeFrancoNemec2026} can go further. For general codes,
space sharing makes the entropic bounds of Grassl, Huber and
Winter~\cite{GHW2022} tight in the capacity sense~\cite{SunJafar2026}.

\section*{Acknowledgments}
\mbox{J.-L.}~Kim was supported in part by the BK21 FOUR (Fostering Outstanding
Universities for Research) funded by the Ministry of Education (MOE, Korea),
National Research Foundation of Korea (NRF) under Grant
No.~4120240415042, Basic Science Research Program through the National
Research Foundation of Korea (NRF) funded by the Ministry of Science and ICT
under Grant No.~RS-2025-24534992 and Global - Learning \& Academic research
institution for Master's$\cdot$PhD students, and Postdocs (LAMP) Program of
the National Research Foundation of Korea (NRF) grant funded by the Ministry
of Education (No.~RS-2024-00441954).

Claude (Anthropic, Opus series) and Codex (OpenAI, GPT-6 series) assisted
with literature search, proofs, programs, figures, the Lean formalization and
the text. The authors checked every result.

\section*{Data availability}
The data that support the findings of this article are openly
available~\cite{DataRecord}. They comprise the generator data for the
codes reported here, the exact evaluators and Magma scripts used to verify
them, the satisfiability instance generators with the CaDiCaL version and
options used, the DRAT certificates, and the Lean~4 project with its
Comparator configurations and verdicts. Every refutation in
Table~\ref{tab:prop13-checks} is regenerated deterministically from them.
Appendix~\ref{sec:machine} states which checks carry certificates.

\appendix
\counterwithin*{equation}{section}
\renewcommand{\theequation}{\thesection\arabic{equation}}

\section{Notation and preliminaries}\label{app:notation}

Throughout, $\Cc$ is the stabilizer of an $\ea{n}{k}{d}{c}_q$ code. Two
facts from Sec.~\ref{sec:prelim} are used, $\dim\Cd=\dim\Hull+2k$ and that
every vector of $\Cd$ outside $\Hull$ has weight at least $d$. A single
qudit is denoted by
$t$ and a set of qudits by $T$. For a set $T$ let $\pi_T$ be the
projection of $\F_q^{2n}$ onto the components at the qudits of $T$, and
we write $\pi_t=\pi_{\{t\}}$. The support $\mathrm{supp}\,\Hull$ is the set of
qudits $t$ with $\pi_t(\Hull)\ne0$.
Table~\ref{tab:appnotation} lists the remaining notation.

\begin{table}[htbp]
\caption{Notation used in the appendices.}
\label{tab:appnotation}
\setlength{\tabcolsep}{3pt}
\centering
\begin{tabular}{lll}
\toprule
Symbol & Meaning & Defined in\\
\hline
$\pi_T$, $\pi_t$ & projection onto $T$, onto $t$ & App.~\ref{app:notation}\\
$\mathrm{supp}\,\Hull$ & qudits $t$ with $\pi_t(\Hull)\ne0$ & App.~\ref{app:notation}\\
$M_q(\ell)$ & largest length of an MDS code of dimension $\ell$ & Sec.~\ref{sec:diag}\\
$e_t$, $e_t'$ & basis of $\F_q^2$, $\pi_t(\Hull)\le\F_qe_t$, $\langle e_t,e_t'\rangle=1$ & App.~\ref{app:diag}\\
$g_t$, $f_t$ & functionals on $\Hull$ and $\Cd/\Hull$ & App.~\ref{app:diag}\\
$N_q(\ell,\epsilon)$ & \begin{tabular}[t]{@{}l@{}}largest length of a code of dimension $\ell$\\and Singleton defect at most $\epsilon$\end{tabular} & Sec.~\ref{sec:further}\\
$B_q(k,\sigma)$ & bound on $\dim\Hull$ built from $N_q$ & Sec.~\ref{sec:further}\\
$\rho_s$ & $Z\otimes Z^{-1}$ on pair $s$ & Sec.~\ref{sec:diagcodes}\\
$L_r$, $p_a$ & logical generators, polynomial of a class & Sec.~\ref{sec:diagcodes}\\
$\eta_s$, $\delta$ & nonzero scalars in the generators & Eq.~\eqref{eq:diagcode}\\
\bottomrule
\end{tabular}
\end{table}

\begin{lemma}[Projections of the hull]\label{lem:perset}
If $|T|\ge n-d+1$, then every vector of $\Cd$ vanishing on $T$ lies in
$\Hull$, and $\dim\pi_T(\Hull)\le2(|T|-k)$.
\end{lemma}

\begin{proof}
A vector vanishing on $T$ has weight at most $n-|T|\le d-1$, so it lies in
$\Hull$. Hence $\pi_T$ has the same kernel on $\Cd$ as on $\Hull$, and
rank--nullity gives $\dim\pi_T(\Cd)-\dim\pi_T(\Hull)=\dim\Cd-\dim\Hull=2k$.
As $\pi_T(\Cd)\le\F_q^{2|T|}$, the bound follows.
\end{proof}

Lemma~\ref{lem:perset} and Theorem~\ref{thm:bdh} use only the two facts
stated above, not the symplectic form.

\section{Proofs}\label{app:proofs}

\begin{figure}[htbp]
\centering
\resizebox{\textwidth}{!}{\input{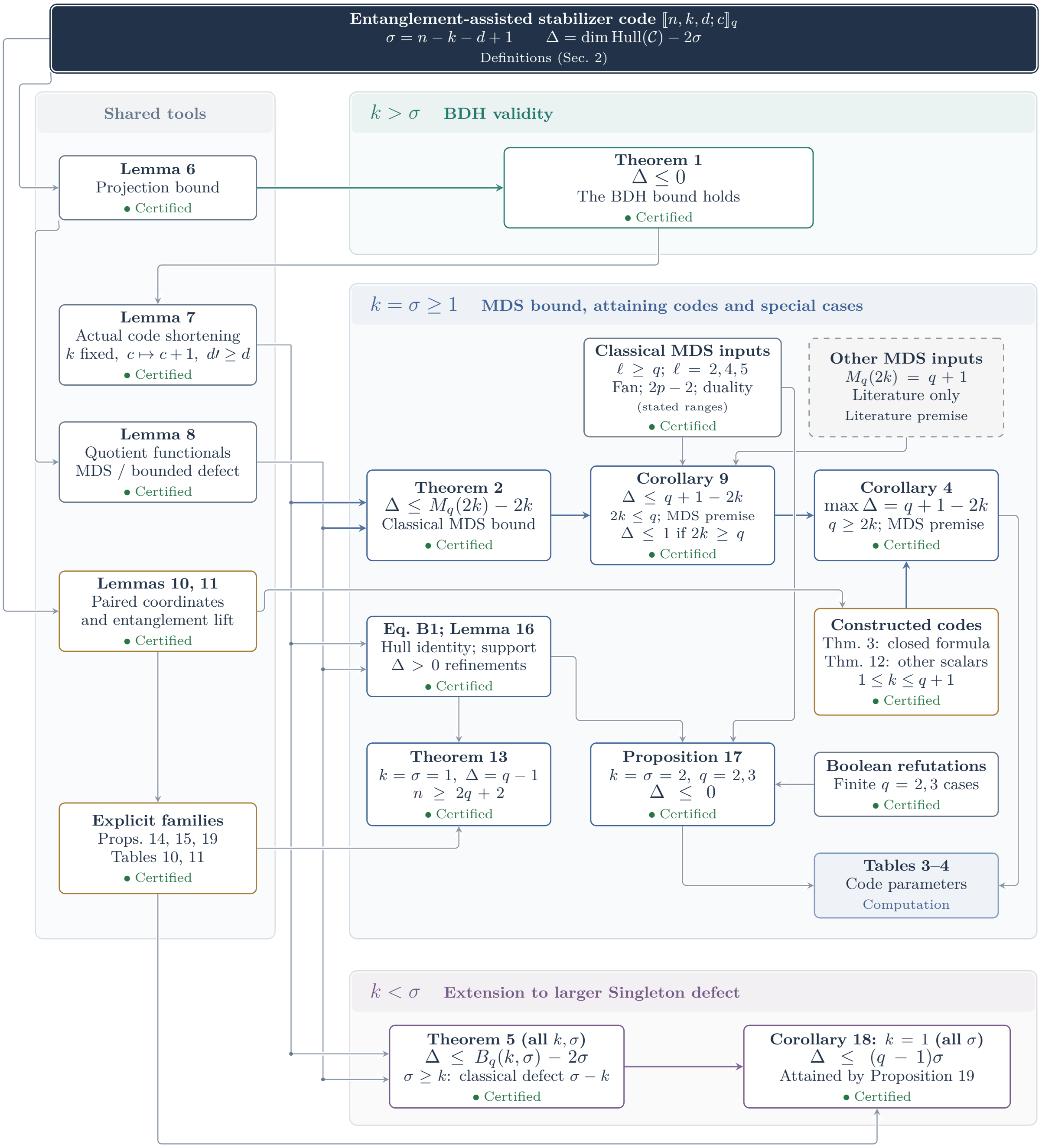}}
\caption{Dependencies among the results and their status in Lean. An arrow
points from a result to one whose proof uses it.}
\label{fig:deps}
\end{figure}

Figure~\ref{fig:deps} shows how the results depend on each other and which
of them are certified in Lean (Appendix~\ref{sec:machine}). An arrow from a
box that groups several results may use only some of them. The shaded bands
group the argument by the three cases $k>\sigma$, $k=\sigma$ and
$k<\sigma$ and are not hypotheses. Theorem~\ref{thm:above} and
Corollary~\ref{cor:k1} hold under the hypotheses stated in their boxes,
which are wider than their band.

\subsection{The case \texorpdfstring{$k>\sigma$}{k > σ}}\label{app:bdh}

\begin{proof}[Proof of Theorem~\ref{thm:bdh}]
Lemma~\ref{lem:perset} at $|T|=n-d+1$ gives $\sigma\ge0$. Suppose
$\dim\Hull\ge2\sigma+1$. Add qudits one at a time to a set $T$, each time
one that raises $\dim\pi_T(\Hull)$. This is possible while
$\dim\pi_T(\Hull)<\dim\Hull$, because the projection onto all qudits is
injective. Each step raises the dimension by at least one, so after at most
$2\sigma+1$ steps $|T|\le2\sigma+1$ and $\dim\pi_T(\Hull)\ge2\sigma+1$. As
$k\ge\sigma+1$, we have $2\sigma+1\le k+\sigma=n-d+1$, so $T$ extends to a
set $T'$ of $n-d+1$ qudits with $\dim\pi_{T'}(\Hull)\ge2\sigma+1$, whereas
Lemma~\ref{lem:perset} gives $\dim\pi_{T'}(\Hull)\le2(n-d+1-k)=2\sigma$.
Hence $\dim\Hull\le2\sigma$, which is $\Delta\le0$ by Eq.~\eqref{eq:excess}.
\end{proof}

\subsection{The case \texorpdfstring{$k=\sigma$}{k = σ}: upper bound on the violation}\label{app:diag}

The proof of Theorem~\ref{thm:bdh} does not apply here, because
$2\sigma+1$ qudits do not fit inside a set of $n-d+1=2k$ qudits. We begin
with the qudits $t$ with $\dim\pi_t(\Hull)=2$.

\begin{lemma}[Shortening]\label{lem:peel}
Let $\pi_t(\Hull)=\F_q^2$ at some qudit $t$. Restricting the vectors of
$\Cd$ that vanish at $t$ to the other $n-1$ qudits gives the symplectic
dual of an $\ea{n-1}{k}{d'}{c+1}_q$ code $\Cc'$ with $d'\ge d$. The hull of
$\Cc'$ is the restriction of the vectors of $\Hull$ that vanish at $t$ and
has dimension $\dim\Hull-2$, and the Singleton defect of $\Cc'$ is at most
$\sigma-1$. Consequently, if $k\ge\sigma$ and $\dim\pi_t(\Hull)=2$ at some
qudit $t$, then $\Delta\le0$.
\end{lemma}

\begin{proof}
As $\pi_t(\Cd)\supseteq\pi_t(\Hull)=\F_q^2$, the vectors of $\Cd$ and of
$\Hull$ that vanish at $t$ form subspaces of dimensions $\dim\Cd-2$ and
$\dim\Hull-2$. Deleting qudit $t$ is injective on them and keeps the
symplectic form and the weights. The quotient of the first subspace by the
second maps injectively, hence isomorphically, onto $\Cd/\Hull$, on which
the form is nondegenerate. So the restricted hull is the hull of the
restricted dual, $k$ is unchanged and $c$ rises by one. A restricted vector
outside the restricted hull has weight at least $d$, so $d'\ge d$ and the
Singleton defect $(n-1)-k-d'+1$ is at most $\sigma-1$. For $k\ge\sigma$ it
is less than $k$, and Theorem~\ref{thm:bdh} gives
$\dim\Hull-2\le2(\sigma-1)$.
\end{proof}

Lemma~\ref{lem:peel} is the propagation rule (7) of Luo
et al.~\cite{Luo2022}. At $k=1$ and $d=n-1$ its last statement is
equivalent to~\cite[Lemma~2]{BaekWonKim2026}.

If $\dim\pi_t(\Hull)\le1$ at every qudit $t$, each qudit in the support of
$\Hull$ carries two linear functionals. Fix $e_t$ spanning $\pi_t(\Hull)$
and $e_t'$ with $\langle e_t,e_t'\rangle=1$ for the form $xz'-zx'$ on
$\F_q^2$, so that $e_t,e_t'$ is a basis. Define $g_t$ on $\Hull$
and $f_t$ on $\Cd$ by $\pi_t(\rho)=g_t(\rho)e_t$ for $\rho\in\Hull$ and
$\pi_t(v)\in f_t(v)e_t'+\pi_t(\Hull)$ for $v\in\Cd$. As $f_t$ vanishes on
$\Hull$, it is a functional on $\Cd/\Hull\cong\F_q^{2k}$. The $g_t$ span the
dual space of $\Hull$, since a vector of $\Hull$ is determined by its
components.

\begin{lemma}[Spanning functionals]\label{lem:arc}
Let $\dim\pi_t(\Hull)\le1$ at every qudit $t$, and let $T$ be a set of
$k+\sigma$ qudits whose $g_t$ are linearly independent. Then the $f_t$,
$t\in T$, span the dual space of $\Cd/\Hull$.
\end{lemma}

\begin{proof}
Otherwise some $v\in\Cd\setminus\Hull$ has $f_t(v)=0$, that is
$\pi_t(v)=a_te_t$ with $a_t\in\F_q$, for every $t\in T$. As the $g_t$ are
independent there is $\rho\in\Hull$ with $g_t(\rho)=a_t$ for $t\in T$, and
then $v-\rho\in\Cd\setminus\Hull$ vanishes on $k+\sigma=n-d+1$ qudits,
which Lemma~\ref{lem:perset} forbids.
\end{proof}

Choose qudits $t_1,\dots,t_j$, $j=\dim\Hull$, whose $g_t$ form a basis of
the dual space of $\Hull$, and consider the classical code
$\{(f_{t_1}(v),\dots,f_{t_j}(v))\mid v\in\Cd\}\le\F_q^{j}$, the code of the
$f_t$. If $j\ge k+\sigma$, Lemma~\ref{lem:arc} applied to every
$(k+\sigma)$-subset says that this code has dimension $2k$ and that no
nonzero codeword vanishes on $k+\sigma$ coordinates, so any $k+\sigma$ of
its $j$ symbols determine the encoded message. It is then a $[j,2k]_q$
code of Singleton defect at most $\sigma-k$, and for $k=\sigma\ge1$ it is
MDS.

\begin{proof}[Proof of Theorem~\ref{thm:diag}]
We show $\dim\Hull\le M_q(2k)$. As $M_q(2k)\ge2k$, the case
$\dim\Hull\le2k$ is immediate. Otherwise $\Delta>0$,
Lemma~\ref{lem:peel} gives $\dim\pi_t(\Hull)\le1$ at every qudit, and the
code of the $f_t$ is an MDS code of length $\dim\Hull$ and dimension $2k$.
\end{proof}

\begin{corollary}[Explicit bounds for $k=\sigma$]\label{cor:diag}
For $k=\sigma\ge1$, every $\ea{n}{k}{d}{c}_q$ code has
\[
 \Delta\le\begin{cases}
 1 & \text{if } 2k\ge q,\\
 q+1-2k & \text{if } 2k\le q \text{ and } M_q(2k)=q+1.
 \end{cases}
\]
In particular $\Delta\le q+1-2k$ for every prime $q\ge2k$.
\end{corollary}

\subsection{The case \texorpdfstring{$k=\sigma$}{k = σ}: codes attaining the bound}\label{app:codes}

Operators $Z\otimes Z^{-1}$ on pairs of qudits, such as $\rho_s$ in
Eq.~\eqref{eq:diagcode}, commute with each other, and $v$ commutes with one of them if and only if its
$x$-components agree on its two qudits.

\begin{lemma}[The operator $Z\otimes Z^{-1}$]\label{lem:pair}
Let $\rho=Z\otimes Z^{-1}$ on two qudits, and let $v$ commute with $\rho$,
with common $x$-value $x_0$ and $z$-components $(z_1,z_2)$ on these two
qudits. Then $v+\nu\rho$ agrees with $v$ off the pair, and the largest number of qudits of the pair outside the support of
$v+\nu\rho$, over $\nu\in\F_q$, is
\[
 \begin{cases}
 0 & \text{if } x_0\ne0,\\
 1 & \text{if } x_0=0 \text{ and } z_1+z_2\ne0,\\
 2 & \text{if } x_0=0 \text{ and } z_1+z_2=0.
 \end{cases}
\]
\end{lemma}

\begin{proof}
The components of $v+\nu\rho$ on the pair are $(x_0,z_1+\nu)$ and
$(x_0,z_2-\nu)$. For $x_0\ne0$ neither vanishes. For $x_0=0$ they vanish at
$\nu=-z_1$ and at $\nu=z_2$, which coincide exactly when
$z_1+z_2=0$.
\end{proof}

\begin{lemma}[Adding one pair]\label{lem:lift}
If an $\ea{n}{k}{d}{c}_q$ code exists with $c\le n-k-1$, then an
$\ea{n}{k}{d'}{c+1}_q$ code exists with $d'\ge d$.
\end{lemma}

\begin{proof}
Pick $\rho\in\Hull$, $\rho\ne0$ (possible since $\dim\Hull=n-k-c\ge1$).
Choose $u\in\F_q^{2n}$ with $\langle\rho,u\rangle=1$ and set
$\Cc'=\langle\Cc,u\rangle$. The Gram rank rises by two, so $c'=c+1$ and
$k'=n-(\dim\Cc+1)+(c+1)=k$. If $v\in\Cc'^{\perp_{\mathrm s}}\setminus\mathrm{Hull}(\Cc')$
then $v\in\Cd$, and $v\notin\Hull$ (else $v\in\Cc\le\Cc'$ with
$v\in\Cc'^{\perp_{\mathrm s}}$ would put $v$ in $\mathrm{Hull}(\Cc')$). Hence
every such $v$ lies in $\Cd\setminus\Hull$, and $d'\ge d$.
\end{proof}

This is the propagation rule (4) of Luo et al.~\cite{Luo2022}.
Appendix~\ref{app:k1ex} uses it for $q=3$, and
Appendix~\ref{sec:machine} uses it to carry codes to more pairs.

\begin{proof}[Proof of Theorem~\ref{thm:diagcodes}]
\emph{(i) Hull.} Each $L_r$ has equal $x$-values on every pair, so it
commutes with every $\rho_s$, and the $\rho_s$ commute with each other.
Hence $\langle\rho_s\rangle\le\Hull$, with equality once the Gram matrix
$G$ of the $L_r$ is shown nondegenerate.

\emph{(ii) Distance.} Let $a\ne0$ and $v=\sum_ra_rL_r$. On a pair $s\in\F_q$ the
common $x$-value of $v$ is $p_a(s)$ and its $z$-components are
$(\eta_sp_a'(s),0)$. On pair $\infty$ they are $a_{2k-1}$ and
$(\delta a_{2k-2},0)$, the value and derivative of $p_a$ at $\infty$ read as
a form of degree $2k-1$. By Lemma~\ref{lem:pair} a vector of the coset $v+\Hull$
has at most $\min(m_s,2)$ zero components on pair $s$, where $m_s$ is the
multiplicity of $s$ as a root, since $p_a(s)=p_a'(s)=0$ exactly for a multiple root,
in every characteristic. Shifts on distinct pairs act independently, and a
nonzero form of degree $2k-1$ has at most $2k-1$ roots on $\mathbb{P}^1$
counted with multiplicity, so every vector of $\Cd\setminus\Hull$ has
weight at least $n-(2k-1)$, with equality for $p_a$ a product of $2k-1$
distinct linear factors.

\emph{(iii) Gram matrix.} On pair $s$ the generators $L_r$ and $L_{r'}$
contribute $s^r\,\eta_sr'\,s^{r'-1}-\eta_sr\,s^{r-1}s^{r'}$, so
$G_{rr'}=(r'-r)\sum_{s\in\F_q}\eta_ss^{r+r'-1}$, plus $\delta$ at
$(r,r')=(2k-1,2k-2)$ and $-\delta$ at $(2k-2,2k-1)$ from pair $\infty$.
Here $\sum_{s\ne0}s^j$ is $-1$ when $(q-1)\mid j$ and $0$ otherwise, with
$j=r+r'-2k+1$ and $|j|\le2k-1\le q-1$. Equality $|j|=q-1$ occurs
only for $q=2k$ and $r=r'$, where the factor $r'-r$ vanishes, so only $j=0$
contributes. The only surviving terms are therefore
$G_{r,2k-1-r}=-(2k-1-2r)$, $G_{01}=-G_{10}=1$ from $s=0$, and the
$\delta$-terms. For $k=1$ the first two cancel and $G_{01}=-\delta\ne0$. For $k\ge2$ the
Pfaffian of $G$ factors into the entries $\pm(2k-1-2r)$, $2\le r\le k-1$,
which are odd integers at most $2k-5$ in absolute value, and the Pfaffian
$(2k-1)(2k-3)-\delta$ of the block on $\{0,1,2k-2,2k-1\}$. Under the stated
condition the first factors are nonzero in $\F_q$, and $\delta\ne(2k-1)(2k-3)$
can be chosen since $q\ge4$.

By (i) and (iii), $\dim\Hull=q+1$, and $\dim\Cd=q+1+2k$, so the code has
$k$ logical qudits and $c=n-k-(q+1)=q+1-k$, and $d=n-2k+1$ by (ii). Then
$\sigma=k$ and $\Delta=(q+1)-2k$.
\end{proof}

\begin{proof}[Proof of Corollary~\ref{cor:diagexact}]
Corollary~\ref{cor:diag} gives $\Delta\le q+1-2k$ for every code with
$k=\sigma$. Theorem~\ref{thm:diagcodes}, or Theorem~\ref{thm:s-full-range}
where its condition on $p$ fails, gives a code with $n=2q+2$ and
$\Delta=q+1-2k$.
\end{proof}

The proof of Theorem~\ref{thm:diagcodes} uses the scalars only through
$\eta_s\ne0$ and $\delta\ne0$ in step (ii) and through $G$ in step
(iii). Where its condition fails,
other nonzero scalars still give these parameters, for every $q$ and
$1\le k\le q+1$.

\begin{theorem}[Every logical dimension]\label{thm:s-full-range}
For every finite field $\F_q$ and every $1\le k\le q+1$, there exist
$\eta_s\in\F_q^\times$ for $s\in\F_q$ and $\delta\in\F_q^\times$ such that
the generators of Eq.~\eqref{eq:diagcode}, with these scalars, define an
\[
 \ea{2q+2}{k}{2q+3-2k}{q+1-k}_q
\]
code.
\end{theorem}

\begin{proof}
We write $w_s=\eta_s$ for $s\in\F_q$ and $w_\infty=\delta$. Step (i) of
the proof of Theorem~\ref{thm:diagcodes} and the lower bound in step (ii)
hold for any nonzero scalars, and they do not use $k\le q/2$. By step (i),
once the Gram matrix $G(w)$ of
the logical generators is nonsingular, $\Hull$ is spanned by the $q+1$
operators $\rho_s$, so the code has $k$ logical qudits and $c=q+1-k$
pairs. By step (ii), every vector of $\Cd\setminus\Hull$ then has weight
at least $2q+3-2k$. It remains to choose $w$ with $G(w)$ nonsingular and to
find a vector of this weight.

Read the polynomial $p_a$ of a logical class $a\in\F_q^{2k}$ as the binary
form
\[
 F_a(X,Y)=\sum_{r=0}^{2k-1}a_rX^rY^{2k-1-r}
\]
of degree $2k-1$. For $s\in\F_q$ let $\varphi_s(F)=F(s,1)$, and let
$\psi_s(F)$ be the derivative of $F(x,1)$ at $x=s$. For $s=\infty$ let
$\varphi_\infty(F)$ and $\psi_\infty(F)$ be the constant and the linear
coefficient of $F(1,y)$. As in step (ii), on pair $s$ the
vector $\sum_ra_rL_r$ has common $x$-value $\varphi_s(F_a)$ and
$z$-components $(w_s\psi_s(F_a),0)$, and $\varphi_s(F)=\psi_s(F)=0$
exactly when $s$ is a root of $F$ of multiplicity at least two.

\emph{Gram matrix.} On pair $s$ the classes $a$ and $a'$ contribute
$w_s\bigl(\varphi_s(F_a)\psi_s(F_{a'})-\psi_s(F_a)\varphi_s(F_{a'})\bigr)$
to the symplectic form. Let $E$ be the $(2q+2)\times2k$ matrix whose rows,
pair by pair, are the coefficient vectors of $\varphi_s$ and $\psi_s$, and
let $J(w)$ be the block-diagonal matrix with the $2\times2$ blocks
$w_s\bigl(\begin{smallmatrix}0&1\\-1&0\end{smallmatrix}\bigr)$. The Gram
matrix of the logical generators is then $G(w)=E^{\mathsf T}J(w)E$.

For a set $I$ of $k$ points let $E_I$ be the $2k\times2k$ matrix formed by
the rows of $E$ that belong to the points of $I$. If $E_Ia=0$, then $F_a$
has a root of multiplicity at least two at every point of $I$, hence at
least $2k$ roots counted with multiplicity, more than its degree. So
$F_a=0$, and $\det E_I\ne0$.

The minor-summation formula of Ishikawa and
Wakayama~\cite{IshikawaWakayama1995} expresses the Pfaffian of
$E^{\mathsf T}AE$, for an alternating matrix $A$, as the sum of
$\operatorname{Pf}(A_R)\det(E_R)$ over all sets $R$ of $2k$ rows, where
$A_R$ and $E_R$ keep only the rows and columns in $R$. For $A=J(w)$ the
matrix $J(w)_R$ has a zero row unless $R$ consists of both rows of each of
$k$ points $I$, and then its Pfaffian is $\prod_{s\in I}w_s$. Hence
\[
 P(w):=\operatorname{Pf}G(w)
   =\sum_{|I|=k}\det(E_I)\prod_{s\in I}w_s.
\]
The formula is an identity between polynomials with integer coefficients,
so it holds over every field, including those of characteristic two.

\emph{Choice of scalars.} Distinct sets $I$ give distinct monomials, and
every coefficient $\det E_I$ is nonzero, so $P$ is a nonzero polynomial of
degree at most one in each $w_s$. Such a polynomial cannot vanish at all
points where each variable takes one of two fixed distinct values. Indeed,
write $P=P_0+w_sP_1$ with $P_0$ and $P_1$ free of $w_s$. Vanishing at both
values of $w_s$ forces $P_0$ and $P_1$ to vanish at all such points of the
other variables, and induction on the number of variables gives $P=0$.
For $q\ge3$ the group $\F_q^\times$ has two distinct elements, so some
$w\in(\F_q^\times)^{q+1}$ has $P(w)\ne0$, and then
$\det G(w)=P(w)^2\ne0$.

For $q=2$ the only nonzero scalar is $1$, and $k\le3$, so we check $G$
directly. At $k=1$ it is
$\bigl(\begin{smallmatrix}0&1\\1&0\end{smallmatrix}\bigr)$, and at $k=2$ it
is the $4\times4$ matrix with ones on the antidiagonal and zeros
elsewhere. At $k=3$ the three points of $\mathbb{P}^1(\F_2)$ form the only
set $I$, so $E=E_I$ is square and invertible, and $G=E^{\mathsf T}JE$ is
nonsingular because $J$ is.

\emph{A vector of least weight.} A logical class $a$ has a vector with
one zero component on the pair of each simple root of $F_a$ and two on the
pair of each multiple root (Sec.~\ref{sec:diagcodes}). The product of
$2k-1$ distinct linear factors used in step (ii) needs $2k-1$ points,
more than $q+1$ when $k$ is large, so we use double roots. For $k\le q$
take a form with $k-1$ distinct finite double roots and one further finite
simple root. It uses $k$ points and gives $2(k-1)+1=2k-1$ zero components.
For $k=q+1$ take
\[
 F(X,Y)=Y(X^q-XY^{q-1})^2
\]
of degree $2q+1=2k-1$. It has a double root at every finite point, because
$x^q-x$ vanishes on $\F_q$, and a simple root at infinity, as
$F(1,y)=y(1-y^{q-1})^2$, which again gives $2k-1$ zero components. Hence
$d=2q+3-2k$ in every case.
\end{proof}

\subsection{The case \texorpdfstring{$k=\sigma$}{k = σ}: one logical qudit and small fields}\label{sec:length}

Here we also use that $\Hull$ is orthogonal to $\Cd$. With $e_t$, $g_t$
and $f_t$ as in Appendix~\ref{app:diag}, for $\rho\in\Hull$ and $v\in\Cd$
the components at $t$ are $g_t(\rho)e_t$ and $f_t(v)e_t'$ plus a multiple
of $e_t$, and $\langle e_t,e_t\rangle=0$, so
$\langle\rho,v\rangle=\sum_tg_t(\rho)f_t(v)$. The orthogonality reads
\begin{equation}\label{eq:radid}
  \sum_{t\in\mathrm{supp}\,\Hull}g_t\otimes f_t=0 .
\end{equation}

\begin{theorem}[One logical qudit]\label{thm:k1length}
(i) Every $\ea{n}{1}{d}{c}_q$ code with $\sigma=1$ and $\Delta=q-1$ has
$|\mathrm{supp}\,\Hull|\ge2q+2$. (ii) $\cmins(2q+1,1,2q)=q$ for every $q$.
(iii) For $q=2$, $\cmins(n,1,n-1)=n-3$ at every odd $n\ge5$.
\end{theorem}

\begin{proof}
(i) At a qudit $t$ of $\mathrm{supp}\,\Hull$ we write
$\pi_t(v)=\mu_t(v)e_t+f_t(v)e_t'$ for $v\in\Cd$, so that $\mu_t$ restricts
to $g_t$ on $\Hull$. As $\sigma=1$ and $\Delta=q-1$,
Eq.~\eqref{eq:excess} gives $\dim\Hull=\Delta+2\sigma=q+1$, and
$\dim\pi_t(\Hull)\le1$ at every qudit by Lemma~\ref{lem:peel}. Choose
qudits $t_1,\dots,t_{q+1}$ whose $g_{t_i}$ form a basis of the dual space
of $\Hull$. By Lemma~\ref{lem:arc} with $k+\sigma=2$ their $f_{t_i}$ are
pairwise independent, so they represent all $q+1$ points of
$\mathrm{PG}(1,q)$. If another qudit $t$ of $\mathrm{supp}\,\Hull$ had
$g_t$ with at least two nonzero coordinates in this basis, $g_t$ and every
$g_{t_i}$ would be independent, and so would $f_t$ and every $f_{t_i}$,
which is impossible as the $f_{t_i}$ exhaust $\mathrm{PG}(1,q)$. Hence
$g_t=\gamma_tg_{t_i}$ and $f_t=\kappa_tf_{t_i}$ for a single $i$. The
$g_{t_i}$-component of Eq.~\eqref{eq:radid} reads
$(1+\sum\gamma_t\kappa_t)f_{t_i}=0$, summed over these $t$, so for every
$i$ there is at least one such $t$, and $|\mathrm{supp}\,\Hull|\ge2q+2$.

(ii) At $n=2q+1$, part (i) and the bound $\Delta\le q-1$ of
Theorem~\ref{thm:diag} leave $\Delta\le q-2$, which the codes of
Proposition~\ref{thm:unified} below attain.

(iii) For $q=2$ the three $f_{t_i}$ are the three nonzero functionals on
$\Cd/\Hull\cong\F_2^2$, and $\gamma_t=\kappa_t=1$. Let $n$ be odd. Here
$d=n-1$, and suppose $\Delta=1$. Given $i$, take $v\in\Cd\setminus\Hull$
with $f_{t_i}(v)=0$. No other basis qudit has $f$ vanishing at $v$, so in
$\mathrm{supp}\,\Hull$ a component of $v+\rho$ ($\rho\in\Hull$) can vanish
only at a qudit $t$ with $g_t=g_{t_i}$, where it is
$(\mu_t(v)+g_{t_i}(\rho))e_t$. One zero component is allowed, so the values
$\mu_t(v)$ at these qudits are distinct. Exactly one qudit $\bar t_i\ne t_i$
has $g_{\bar t_i}=g_{t_i}$, with $\mu_{t_i}(v)\ne\mu_{\bar t_i}(v)$, and every class of $\Cd/\Hull$ contains a
vector with a zero component in $\mathrm{supp}\,\Hull$. A qudit outside
$\mathrm{supp}\,\Hull$ must then map $\Cd/\Hull$ bijectively onto
$\F_2^2$, adding the nonzero alternating form $\omega_0$ of $\Cd/\Hull$ to
the form that $\langle\cdot,\cdot\rangle$ induces on $\Cd/\Hull$. The pair
$t_i,\bar t_i$ adds $(\mu_{t_i}+\mu_{\bar t_i})\wedge f_{t_i}$, where
$\mu_{t_i}+\mu_{\bar t_i}$ vanishes on $\Hull$ but not at $v$. This is $\omega_0$
as well. The induced form is therefore $(n-3)\,\omega_0=0$. But this form
is nondegenerate, because $\Hull$ is the radical of the form on $\Cd$, a
contradiction. Hence $\cmins(n,1,n-1)\ge n-3$, and the codes of
Proposition~\ref{thm:unified} attain it.
\end{proof}

At every $n\ge2q+2$, Theorem~\ref{thm:diag} gives $\Delta\le q-1$, and the
codes of Proposition~\ref{thm:floorfam} below attain it, except for $q=2$
and odd $n$, where part (iii) applies. With part (ii) this determines
$\cmins(n,1,n-1)$ at every $n\ge2q+1$.

The codes of Propositions~\ref{thm:floorfam} and~\ref{thm:unified}, like
those of Eq.~\eqref{eq:diagcode}, are given by logical generators and the
operators $Z\otimes Z^{-1}$ on disjoint pairs of qudits, so by Lemma~\ref{lem:pair} the least weight in a logical class is
found pair by pair. In Tables~\ref{tab:s-floor} and~\ref{tab:s-frame} the
components $(x,z)$ of consecutive qudits are written side by side.

Fix a prime power $q=p^h$ and $n\ge 2q+2$. Let pairs $0,\dots,q$ be the
qudits $\{1,2\},\dots,\{2q+1,2q+2\}$, with $\rho_i=Z\otimes Z^{-1}$ on pair
$i$, and call the other $n-2q-2$ qudits the remaining qudits. Let
$u_1,\dots,u_q$ enumerate $\F_q$, and choose $u_0\in\F_q^{\times}$. Define
$a$ and $b$ by their components $(x,z)$ in Table~\ref{tab:s-floor}, and let $\Cc$
be the code with $\Cd=\langle a,b,\rho_0,\dots,\rho_q\rangle$.
\begin{table}[htbp]
\caption{Components $(x,z)$ of $a$ and $b$ for the family with $c=n-q-2$.}
\label{tab:s-floor}
\centering
\begin{tabular}{cccc}
\toprule
 & Pair $0$ & Pair $i\ge1$ & Remaining qudit\\
\hline
$a$ & $(0,0)(0,1)$ & $(1,0)(1,0)$ & $(0,1)$\\
$b$ & $(u_0,0)(u_0,1)$ & $(u_i,0)(u_i,1)$ & $(1,0)$\\
\bottomrule
\end{tabular}
\end{table}
As Pauli operators, $a$ is $I\otimes Z$ on pair $0$, $X\otimes X$ on pairs
$1,\dots,q$ and $Z$ on each remaining qudit, and $b$ is
$X^{u_i}\otimes X^{u_i}Z$ on pair $i$ and $X$ on each remaining qudit.

\begin{proposition}[Family with $c=n-q-2$]\label{thm:floorfam}
Let $q$ be a prime power and $n\ge2q+2$. If $u_0\in\F_q^{\times}$ satisfies
$u_0+n-2\ne0$ in $\F_q$, then the code $\Cc$ of Table~\ref{tab:s-floor} is
an $\ea{n}{1}{n-1}{n-q-2}_q$ code, with $\Delta=q-1$. Such a $u_0$ exists
for every $q\ge3$ and every $n\ge2q+2$, and for $q=2$ exactly when $n$ is
even.
\end{proposition}

\begin{proof}
Both $a$ and $b$ have equal $x$-values on the two qudits of every pair,
so they commute with every $\rho_i$, and the $\rho_i$ commute with each
other (Appendix~\ref{app:codes}). Summing over
qudits, $\langle a,b\rangle=-u_0+q-(n-2q-2)$, computed in $\F_q$. Since $p$
divides $q$,
\[
  \langle a,b\rangle=-(u_0+n-2),
\]
nonzero exactly under the stated condition. Thus the form on the span of
$a,b$ is nondegenerate, and $\Hull=\langle\rho_0,\dots,\rho_q\rangle$ has dimension
$q+1$.

Let $v=\alpha a+\beta b+\sum_i \nu_i\rho_i$ with $(\alpha,\beta)\ne(0,0)$. On
a remaining qudit $v$ has component $(\beta,\alpha)\ne(0,0)$, so the remaining
qudits contribute $n-2q-2$. On
pair $0$ the components are $(\beta u_0,\,\nu_0)$ and
$(\beta u_0,\,\alpha+\beta-\nu_0)$. If $\beta\ne0$ both are nonzero,
contributing $2$. If $\beta=0$ then $\alpha\ne0$, and $\nu_0=0$ makes the
component on the first qudit zero, contributing $1$ and never $0$. On pair $i\ge1$ the components are
$(\alpha+\beta u_i,\,\nu_i)$ and $(\alpha+\beta u_i,\,\beta-\nu_i)$,
contributing $2$ when $\alpha+\beta u_i\ne0$ and at least $1$ when it
vanishes. In the latter case $\beta\ne0$, and choosing $\nu_i=0$ makes the
contribution exactly $1$, and it can never be $0$.

If $\beta=0$, only pair $0$ can contribute less than $2$, and its minimum
contribution is $1$. If $\beta\ne0$, pair $0$ contributes $2$ and
$\alpha+\beta u_i$ vanishes for exactly one $i\ge1$, because
$u_1,\dots,u_q$ exhaust $\F_q$. Thus every vector in a nonzero logical
class has weight at least $(n-2q-2)+1+2q=n-1$, and a suitable multiple of
$\rho_i$ on the unique pair with zero $x$-value attains this weight in each
such class.
Finally $\dim\Cd=q+3$, so $c=n-1-(q+1)=n-q-2$,
$k=1$ and $d=n-1$, whence $\Delta=2(n-2)-(n-1+n-q-2)=q-1$.
\end{proof}

We repeat the construction of Theorem~1 of our earlier
work~\cite{BaekWonKim2026} and its proof in the notation used here. Fix a prime power $q=p^h$ and
$n\ge 2q+1$. Let pairs $1,\dots,q$ be the qudits
$\{1,2\},\dots,\{2q-1,2q\}$, with $\rho_i=Z\otimes Z^{-1}$ on pair $i$,
and let $u_1,\dots,u_q$ enumerate $\F_q$. Define $a$ and $b$ by their
components $(x,z)$ in Table~\ref{tab:s-frame}, where $\zeta=1$ if $p$ is
odd and $\zeta=0$ if $p=2$, and $\lambda\in\F_q^{\times}$. Let $\Cc$ be the
code with $\Cd=\langle a,b,\rho_1,\dots,\rho_q\rangle$.
\begin{table}[htbp]
\caption{Components $(x,z)$ of $a$ and $b$ for the family with $c=n-q-1$.}
\label{tab:s-frame}
\centering
\begin{tabular}{cccc}
\toprule
 & Pair $i$ & Qudits $2q+1,\dots,n-1$ & Qudit $n$\\
\hline
$a$ & $(1,0)(1,0)$ & $(1,0)$ & $(1,0)$\\
$b$ & $(u_i,1)(u_i,\zeta)$ & $(0,1)$ & $(0,\lambda)$\\
\bottomrule
\end{tabular}
\end{table}
As Pauli operators, $a=X^{\otimes n}$, and $b$ is
$X^{u_i}Z\otimes X^{u_i}Z^{\zeta}$ on pair $i$, $Z$ on qudits
$2q+1,\dots,n-1$ and $Z^{\lambda}$ on qudit $n$.

\begin{proposition}[Family with $c=n-q-1$]\label{thm:unified}
Let $n\ge2q+1$, and let $q\ge3$, or $q=2$ and $n$ odd. Then for a suitable
$\lambda$ the code $\Cc$ of Table~\ref{tab:s-frame} is an
$\ea{n}{1}{n-1}{n-q-1}_q$ code, with $\Delta=q-2$. When $q=2$ and $n$ is
even no $\lambda$ succeeds.
\end{proposition}

\begin{proof}
On pair $i$ the vector $a$ has $x$-values $1,1$ and $b$ has $u_i,u_i$, so
both commute with $\rho_i$, and the $\rho_i$ commute with each other. As
$a$ has component $(1,0)$ at every qudit, $\langle a,b\rangle$ is the sum
of the $z$-components of $b$,
and since $p$ divides $q$
\begin{align*}
  \langle a,b\rangle&= q\,(1+\zeta)+(n-2q-1)+\lambda\\
  &=n-1+\lambda ,
\end{align*}
computed in $\F_q$. So $\langle a,b\rangle\ne0$ precisely when
$\lambda\ne-(n-1)$ in $\F_q$. A choice exists when $q\ge3$, while for $q=2$
the only $\lambda$ is $1$ and $\langle a,b\rangle=n\bmod 2$. Then
$\Hull=\langle\rho_1,\dots,\rho_q\rangle$, of dimension $q$.

Let $v=\alpha a+\beta b$ with $(\alpha,\beta)\ne(0,0)$ and
$\rho=\sum_i \nu_i\rho_i$. On qudits $2q+1,\dots,n$ the components of $v+\rho$ are $(\alpha,\beta)$ or
$(\alpha,\lambda\beta)$, all nonzero. If $\beta=0$, $\alpha\ne0$ is the
$x$-value on every pair qudit and $\wt(v+\rho)=n$. If $\beta\ne0$, the
$x$-value $\alpha+\beta u_i$ vanishes on exactly one pair, where the components
are $(0,\beta+\nu_i)$ and $(0,\beta\zeta-\nu_i)$. By Lemma~\ref{lem:pair}
at most one of them vanishes, because $\beta(1+\zeta)\ne0$ in both
characteristics. Hence every
vector of $\Cd\setminus\Hull$ has weight at least $n-1$, with equality
attained. Finally $\dim\Cd=q+2$, so $k=1$, $c=n-1-q$, $d=n-1$ and
$\Delta=2(n-2)-(n-1+c)=q-2$.
\end{proof}

\begin{lemma}[Support of the hull]\label{lem:support}
Every $\ea{n}{k}{d}{c}_q$ code with $k=\sigma$ and $\Delta>0$ has
$|\mathrm{supp}\,\Hull|\ge\dim\Hull+2k$.
\end{lemma}

\begin{proof}
By Lemma~\ref{lem:peel}, $\dim\pi_t(\Hull)\le1$ at every qudit. The $g_t$
span the dual space of $\Hull$, and as $\dim\Hull>2k=k+\sigma$ some $2k$ of
them are independent, so by Lemma~\ref{lem:arc} the $f_t$ span the dual
space of $\Cd/\Hull$. By~\eqref{eq:radid} the columns of the matrix of the
$f_t$, of rank $2k$, lie in the kernel of the transpose of the matrix of
the $g_t$, of rank $\dim\Hull$.
\end{proof}

\begin{proposition}[Fields of order two and three]\label{prop:k2}
For $q\in\{2,3\}$, every $\ea{n}{2}{d}{c}_q$ code with $\sigma=2$ has
$\Delta\le0$.
\end{proposition}

\begin{proof}[Proof (computer assisted)]
By Corollary~\ref{cor:diag} ($2k=4\ge q$), $\Delta\le1$. Suppose
$\Delta=1$, so $\dim\Hull=5$ and $n-d=3$. By
Lemma~\ref{lem:support}, $|\mathrm{supp}\,\Hull|\ge9$. Average over
$\rho\in\Hull$ the number of zero components of $v+\rho$ and sum over the
$q^4-1$ nonzero classes $v$ of $\Cd/\Hull$. Each qudit of
$\mathrm{supp}\,\Hull$ contributes $(q^3-1)/q$, each other qudit at least
$q^2-1$, and the total is at most $3(q^4-1)$. Hence
$|\mathrm{supp}\,\Hull|\le12$ for $q=2$ and $\le27$ for $q=3$. Qudits
outside $\mathrm{supp}\,\Hull$ do not enter~\eqref{eq:radid} and only add
zero components, so the qudits of $\mathrm{supp}\,\Hull$ alone satisfy the
weight condition. The conditions derived below are necessary for a code
with $\Delta=1$, not sufficient, so it suffices to show that they cannot be
met. Take a set $T$ of five of them whose $g_t$ form a basis
of the dual space of $\Hull$.
By Lemma~\ref{lem:arc} any four of their $f_t$ are independent, so they form
a frame of $\mathrm{PG}(3,q)$, which for $q\le3$ is an arc of the largest
size $M_q(4)=5$. For every other qudit, $g_t$ therefore has at most three
nonzero coordinates in that basis, and $f_t$ is not in the plane through
the $f_{t'}$, $t'\in T'$, for any three qudits $T'\subset T$ that do not
contain all nonzero coordinates of $g_t$. With the frame mapped to the
standard one, each value of $|\mathrm{supp}\,\Hull|$ leaves a Boolean
satisfiability problem in the unknown $g_t$ and $f_t$ of the other qudits.
The solver CaDiCaL~\cite{CaDiCaL} finds every one of them unsatisfiable
(Table~\ref{tab:prop13-checks}) and writes a proof of unsatisfiability in
the DRAT format, which the checker \texttt{drat-trim}~\cite{DRATtrim}
verifies. The conclusion therefore does not rest on the solver.
\end{proof}

Table~\ref{tab:prop13-checks} lists the refuted instances, which have
$k=\sigma=2$, $\Delta=1$ and the given hull support, and the code
$\ea{8}{2}{5}{2}_3$ of Sec.~\ref{sec:small}, which attains $\Delta=0$.
Appendix~\ref{sec:machine} describes the encoding and its checks.

\begin{table}[htbp]
\caption{Instances for Proposition~\ref{prop:k2}.}
\label{tab:prop13-checks}
\setlength{\tabcolsep}{3pt}
\centering
\begin{tabular}{llll}
\toprule
$q$ & Instance & Result & Established by\\
\hline
$2$ & $9\le|\mathrm{supp}\,\Hull|\le12$ & unsatisfiable & DRAT, Lean\\
$3$ & $9\le|\mathrm{supp}\,\Hull|\le27$ & unsatisfiable & DRAT, Lean\\
 & $\ea{8}{2}{5}{2}_{3}$ & $\Delta=0$ & exhaustive\\
\bottomrule
\end{tabular}
\end{table}

\subsection{The case \texorpdfstring{$k<\sigma$}{k < σ}}\label{sec:above}

\begin{proof}[Proof of Theorem~\ref{thm:above}]
For $\sigma<k$ this is Theorem~\ref{thm:bdh}. For $\sigma\ge k$ and
$\dim\pi_t(\Hull)\le1$ at every qudit, the code of the $f_t$ has Singleton
defect at most $\sigma-k$ when $\dim\Hull\ge k+\sigma$, and
$k+\sigma\le N_q(2k,\sigma-k)$ in any case. Otherwise
Lemma~\ref{lem:peel} gives a code with Singleton defect $\sigma'\le\sigma-1$
and hull dimension $\dim\Hull-2$. By induction on $\sigma$,
$\dim\Hull-2\le B_q(k,\sigma')\le B_q(k,\sigma-1)$, as $B_q(k,\sigma)$ is
nondecreasing in $\sigma$.
\end{proof}

At $k=1$ the Griesmer bound gives $N_q(2,\epsilon)=(q+1)(\epsilon+1)$,
attained by repeating every point of $\mathbb{P}^1(\F_q)$, so
$B_q(1,\sigma)=(q+1)\sigma$.

\begin{corollary}[Exact bound for $k=1$]\label{cor:k1}
Every $\ea{n}{1}{d}{c}_q$ code has $\Delta\le(q-1)\sigma$, and for every
$q$ and $\sigma\ge1$ some $\ea{n}{1}{d}{c}_q$ code has
$\Delta=(q-1)\sigma$.
\end{corollary}

\begin{proof}
The bound is Theorem~\ref{thm:above} with $B_q(1,\sigma)=(q+1)\sigma$ and
Eq.~\eqref{eq:excess}. Proposition~\ref{prop:k1codes} below gives
codes that attain it for every $q$ and $\sigma\ge1$.
\end{proof}

The codes that attain Corollary~\ref{cor:k1} follow the classical
construction above and place $\sigma$ pairs over each point of
$\mathbb{P}^1(\F_q)=\F_q\cup\{\infty\}$. Fix $q=p^h$ and $\sigma\ge1$, and
let $\tau=1$ if $q=2$ and $\sigma$ is even, and $\tau=0$ otherwise. For
each $s\in\mathbb{P}^1(\F_q)$ and $j=1,\dots,\sigma$ take a pair $(s,j)$ of
qudits, with $\rho_{s,j}$ the operator $Z\otimes Z^{-1}$ on it, and if
$\tau=1$ one further qudit, so that $n=2\sigma(q+1)+\tau$. Choose
$\gamma_1,\dots,\gamma_\sigma\in\F_q^{\times}$, and let $a$ and $b$ act on
the two qudits of pair $(s,j)$ and on the further qudit as
\begin{equation}\label{eq:k1codes}
\begin{array}{c@{\quad}ccc}
 & s\in\F_q & s=\infty & \text{Further qudit}\\[2pt]
\hline
\rule{0pt}{11pt}a & X\otimes X & Z^{\gamma_j}\otimes I & X\\
b & X^sZ^{\gamma_j}\otimes X^s & X\otimes X & Z
\end{array}
\end{equation}
Let $\Cc$ be the symplectic dual of $\langle a,b,\rho_{s,j}\rangle$, so
that $\Cd$ is this span.

\begin{proposition}[Codes attaining the bound for $k=1$]\label{prop:k1codes}
If $\gamma_1+\dots+\gamma_\sigma\ne\tau$ in $\F_q$, then the code $\Cc$
of~\eqref{eq:k1codes} is an
$\ea{n}{1}{n-\sigma}{\sigma(q+1)-1+\tau}_q$ code with
$\Delta=(q-1)\sigma$. Such scalars exist for every $q$ and $\sigma\ge1$.
\end{proposition}

\begin{proof}
Both $a$ and $b$ have equal $x$-components on the two qudits of every
pair, so they commute with every $\rho_{s,j}$. In $\langle a,b\rangle$
each pair over $s\in\F_q$ contributes $\gamma_j$, each pair over $\infty$
contributes $-\gamma_j$, and the further qudit contributes $\tau$. As $p$
divides $q$, $\langle a,b\rangle=\tau-(\gamma_1+\dots+\gamma_\sigma)\ne0$.
Hence $\Hull$ is spanned by the $\sigma(q+1)$ operators $\rho_{s,j}$, so
$k=1$ and $c=n-1-\sigma(q+1)$.

Let $v=\alpha a+\beta b$ with $(\alpha,\beta)\ne(0,0)$. Its common
$x$-value is $\alpha+\beta s$ on the pairs over $s\in\F_q$ and $\beta$ on
the pairs over $\infty$, so it vanishes over exactly one point $s_0$. On
each pair $(s_0,j)$ the $z$-components of $v$ are $(\beta\gamma_j,0)$ if
$s_0\in\F_q$, where $\beta\ne0$, and $(\alpha\gamma_j,0)$ if $s_0=\infty$,
where $\alpha\ne0$. In both cases their sum is nonzero. By
Lemma~\ref{lem:pair}, a vector of the coset $v+\Hull$ has at most one zero
component on each of the $\sigma$ pairs over $s_0$ and none on the other
pairs, and some vector of the coset has one on each of these $\sigma$ pairs.
The further qudit, if present, has the nonzero component $(\alpha,\beta)$.
Hence $d=n-\sigma$, the Singleton defect is $\sigma$, and
$\Delta=\dim\Hull-2\sigma=(q-1)\sigma$ by Eq.~\eqref{eq:excess}.

For $q\ge3$, where $\tau=0$, take every $\gamma_j=1$ if $p$ does not
divide $\sigma$, and otherwise replace $\gamma_1$ by an element other than
$0$ and $1$. For $q=2$ every $\gamma_j=1$, and $\sigma\ne\tau$ in $\F_2$ by
the choice of $\tau$.
\end{proof}

The scalars must be allowed to differ between the copies $j$. With one
scalar $\gamma$ for all of them the sum is $\sigma\gamma$, which vanishes
whenever $p$ divides $\sigma$.

For $2\le k<\sigma$ the values of $N_q$ are known only in
tables, and whether $B_q(k,\sigma)$ is attained is open.

\section{Examples and comparisons}\label{app:examples}

\subsection{The case \texorpdfstring{$k>\sigma$}{k > σ}}\label{app:bdh-ex}

The hypothesis $k>\sigma$ is needed only for codes that are not
$\F_{q^2}$-linear. If $\Cd$ and $\Hull$ are $\F_{q^2}$-linear, with each
qudit space $\F_q^2$ read as $\F_{q^2}$, a qudit raises the
$\F_{q^2}$-dimension of $\pi_T(\Hull)$ by at most one, $\sigma+1$ qudits
suffice in the proof of Theorem~\ref{thm:bdh}, and the conclusion holds for every $k\ge1$. It is then the
relative Singleton bound for a pair of nested linear codes of Luo,
Mitrpant, Han Vinck and Chen~\cite{LuoRGHW2005}. For codes that are only
$\F_q$-linear a qudit can raise the dimension by two, and the condition
cannot be dropped (Sec.~\ref{sec:bdh}).

Theorem~\ref{thm:bdh} needs $\F_q$-linearity, not only additivity. Run Grassl's
$\ea{10}{2}{7}{2}_{11}$ code and an $\ea{10}{4}{7}{6}_{11}$ code, built from
a $[10,4,7]_{11}$ code $C$ with $C\cap C^{\perp}=0$, on the two
$11$-dimensional factors of each of ten $121$-dimensional qudits. The result
is an additive $\eaa{10}{121^3}{7}{4}_{121}$ code, $\F_{11}$-linear but not
$\F_{121}$-linear, with $k=3>\sigma=1$ and $\Delta=1$.

The refined Singleton bound of Lai and
Ashikhmin~\cite[Theorem~7]{LaiAshikhmin2018} applies for $d>(n+2)/2$. They
evaluated it at $n=9$ and $c=1$, where it gives $2^k\le3.73$, $5.63$ and
$5.18$ for $d=6$, $7$ and~$8$. Table~\ref{tab:s-la} lists the largest $k$
that this bound, Eq.~\eqref{eq:ghw} and our results allow there. Our
results cover every qubit EA stabilizer code (Sec.~\ref{sec:bdh}), and the
bound of Lai and Ashikhmin also nonadditive codes. A code
$\ea{9}{1}{6}{1}_2$ exists~\cite{LaiAshikhmin2018}.

\begin{table}[htbp]
\caption{Largest $k$ allowed at $n=9$, $c=1$, $q=2$.}
\label{tab:s-la}
\centering
\begin{tabular}{lllll}
\toprule
$d$ & Lai--Ashikhmin & Eq.~\eqref{eq:ghw} & Ours & Larger $k$ excluded by\\
\hline
$6$ & $1$ & $1$ & $1$ & Prop.~\ref{prop:k2}, Thm.~\ref{thm:bdh}\\
$7$ & $2$ & $1$ & $0$ & Cor.~\ref{cor:k1}, Thm.~\ref{thm:bdh}\\
$8$ & $2$ & $1$ & $0$ & Thm.~\ref{thm:diag}, Thm.~\ref{thm:bdh}\\
\bottomrule
\end{tabular}
\end{table}

\subsection{Known values of \texorpdfstring{$M_q(\ell)$}{Mq(l)}}\label{app:mq}

The values of $M_q(\ell)$ that matter here are known~\cite{BallLavrauw2019}.
Trivially $M_q(2)=q+1$, and $M_q(\ell)=\ell+1$ whenever $\ell\ge q$.
The table below lists the further cases in which $M_q(\ell)=q+1$ is known.
For $\ell\ge2$ the columns of a generator matrix of an MDS $[m,\ell]_q$ code
form an arc of $m$ points in $\mathrm{PG}(\ell-1,q)$, and every such arc
arises in this way~\cite{BallLavrauw2019}. An arc of size $q+2$ exists in $\mathrm{PG}(\ell-1,q)$ exactly when one
exists in $\mathrm{PG}(q+1-\ell,q)$~\cite[Cor.~19]{BallLavrauw2019}, and a
planar arc has at most $q+1$ points for odd
$q$~\cite[Thm.~45]{BallLavrauw2019}. By these two facts each range of
$\ell$ in the table gives the dual range in its row. The cases $\ell=6$
and $\ell\le2p-2$ are due to Casse and Glynn and to Ball and De Beule.

\begin{center}
\small
\setlength{\tabcolsep}{2.5pt}
\medmuskip=2mu plus 1mu minus 1mu
\thickmuskip=3mu plus 2mu
\begin{tabular}{@{}llll@{}}
\hline\hline
$\ell$ & Dual range & $q$ & Source\\
\hline
$4$ & $q-2$ & odd $q\ge5$ & Segre~\cite{Segre1955}\\
$4$ & $q-2$ & even $q\ge4$ & Casse~\cite{Casse1969}\\
$5$ & $q-3$ & $q\ge5$ & \cite[Thm.~63]{BallLavrauw2019}\\
$6$ & $q-4$ & even $q\ge8$ & \cite[Sec.~19]{BallLavrauw2019}\\
$4\le\ell\le p$ & $q+2-p\le\ell\le q-1$ & $q=p^h$ & Ball~\cite{Ball2012}\\
$\ell\le2p-2$ & $q+4-2p\le\ell\le q-1$ & $q$ not prime & \cite[Thm.~35]{BallLavrauw2019}\\
$2\le\ell\le\beta$ & $q+2-\beta\le\ell\le q$ & odd $q$ & Fan~\cite{Fan2026}\\
$4\le\ell\le q-2$ & & $q\le27$ & \cite[Table~3.7]{HS2001}\\
\hline\hline
\end{tabular}
\end{center}

Here $\beta=\lfloor((p-2)q+6p-10)/(2p-3)\rfloor$. The range of Ball covers
every prime $q\ge\ell$, and the ranges of Fan cover $q=49$.
Appendix~\ref{sec:machine} states which of these ranges are formalized. For
$q=27$ and $12\le\ell\le16$ the last row rests on computations of Chao and
Kaneta reported in~\cite{HS2001}. Hence $M_q(2k)=q+1$ for every $2k\le q\le31$. Without further
input the first open case is $q=32$ with $8\le2k\le26$. Chen and
Huang~\cite{ChenHuang2026} and, independently, Kumar~\cite{Kumar2026} have
announced that every linear MDS code over a
field of even order $q$ has length at most $q+1$ unless its dimension is
$3$ or $q-1$. As $2k$ is neither, this gives $M_q(2k)=q+1$ for every even
$q$ and $2k\le q$. Together with the
results above, $M_q(2k)=q+1$ for every
$2k\le q$ when $q$ is prime, even, or less than $81$, and the first open
cases are $q=81$ with $30\le2k\le52$, $q=121$ with $2k=62$, and $q=125$
with $58\le2k\le70$.

\subsection{The case \texorpdfstring{$k=\sigma$}{k = σ}: codes attaining the bound}\label{app:codes-ex}

Evaluation on $\mathbb{P}^1(\F_q)$ also underlies EA codes from projective
Reed--Solomon codes~\cite{GimenezRuano2023}. What is particular to the
codes~\eqref{eq:diagcode} is the derivative on the first qudit of each pair,
which gives one zero component at a simple root and two at a multiple root.
In the language of Grassl~\cite{Grassl2021}, the $x$-parts form the doubled
projective Reed--Solomon code $C$ and the hull is cut from $(0|C^{\perp})$
down to the span of the $\rho_s$. With all of $(0|C^{\perp})$ as hull the same
generators give a teleportation-type code with $c=k$ and distance at least
$2q+4-4k$, the distance of $C$.

Where the condition of Theorem~\ref{thm:diagcodes} fails with $q\le32$,
the codes of Eq.~\eqref{eq:diagcode} are taken with the scalars in
Table~\ref{tab:s-scalars}. An element of $\F_q$, $q=p^h$, is written as
the integer $\sum_ie_ip^i$ ($0\le e_i<p$) standing for
$\sum_ie_i\theta^i$, where $\theta$ is a root of the Conway polynomial
($x^2+2x+2$ for $q=9$, $x^2+4x+2$ for $q=25$, $x^3+2x+1$ for $q=27$), and
the pair of a point $s\in\F_q$ is indexed by the same integer. For each
$q$ one choice works for every $k$ concerned. With it the Gram matrix of
the $2k$ logical generators has rank $2k$, in our arithmetic and in Magma,
so by steps (i) and (ii) of the proof of Theorem~\ref{thm:diagcodes} these
codes have the parameters stated there.

\begin{table}[htbp]
\caption{Scalars of the codes of Eq.~\eqref{eq:diagcode} found by search.}
\label{tab:s-scalars}
\centering
\setlength{\tabcolsep}{5pt}
\begin{tabular}{*{13}{c}}
\hline\hline
$q$ & $k$ & $\delta$ & $s$ & $\eta_s$ & $\eta_{s+1}$ & $\eta_{s+2}$ & $\eta_{s+3}$ & $\eta_{s+4}$ & $\eta_{s+5}$ & $\eta_{s+6}$ & $\eta_{s+7}$ & $\eta_{s+8}$\\
\hline
$9$ & $4$ & $2$ & $0$ & $3$ & $2$ & $5$ & $2$ & $8$ & $8$ & $8$ & $7$ & $4$\\
\hline
$25$ & $5$--$12$ & $11$ & $0$ & $5$ & $19$ & $3$ & $9$ & $4$ & $16$ & $15$ & $16$ & $21$\\
& & & $9$ & $13$ & $7$ & $4$ & $16$ & $1$ & $13$ & $14$ & $20$ & $1$\\
& & & $18$ & $23$ & $15$ & $9$ & $24$ & $8$ & $19$ & $4$ &  & \\
\hline
$27$ & $4$--$13$ & $24$ & $0$ & $5$ & $19$ & $26$ & $25$ & $3$ & $9$ & $4$ & $16$ & $25$\\
& & & $9$ & $15$ & $16$ & $21$ & $13$ & $26$ & $7$ & $4$ & $16$ & $1$\\
& & & $18$ & $13$ & $14$ & $20$ & $25$ & $25$ & $1$ & $23$ & $15$ & $9$\\
\hline\hline
\end{tabular}
\end{table}

Beyond $k\le q/2$, for $q=9$ and $k=5$ one choice is $\delta=7$ and
\[
 (\eta_0,\ldots,\eta_8)=(6,3,3,3,5,2,2,8,4)
\]
in the same convention. The choice $(\eta_0,\eta_1,\eta_2)=(1,1,2)$,
$\delta=1$, gives the code $\ea{8}{2}{5}{2}_3$ of
Table~\ref{tab:prop13-checks}.

\subsection{The case \texorpdfstring{$k=\sigma$}{k = σ}: one logical qudit}\label{app:k1ex}

The codes of Proposition~\ref{thm:floorfam} are distance-optimal by Eq.~\eqref{eq:ghw}, which at $k=1$
excludes $d=n$ below $c=n-1$. Codes with these parameters are among the
code records released with~\cite{BaekWonKim2026} for $q=2$ at
$n=6,8,10,12$, for $q=3$ at $8\le n\le15$, for $q=4$ at $n=10,12,14$ and
for $q=5$ at $n=12,14$.

Binary $\ea{n}{1}{n-1}{n-3}_2$ codes
for every $n\ge4$ are also constructed by Li, Ren, Guan and
Liu~\cite{LiRenGuanLiu2026}. The lower bound $c\ge n-3$ at odd $n$ is what
Theorem~\ref{thm:k1length} adds.

At $q=3$ one application of Lemma~\ref{lem:lift} to the codes of
Proposition~\ref{thm:unified} gives
$\ea{n}{1}{d'}{n-3}_3$ codes with $d'\ge n-1$, and $d'=n$ is excluded by
Eq.~\eqref{eq:ghw}. Each of these codes is distance-optimal by
Eq.~\eqref{eq:ghw}.
Appendix~\ref{sec:machine} compares these codes with the tables of
codetables.de.

\subsection{The case \texorpdfstring{$k<\sigma$}{k < σ}}\label{app:above-ex}

At $(k,\sigma,q)=(1,2,2)$ Corollary~\ref{cor:k1} excludes $\ea{9}{1}{7}{1}_2$, although these
parameters satisfy~\eqref{eq:ghw}.

For $\sigma=1$ the codes of Proposition~\ref{prop:k1codes} have the
parameters $\ea{2q+2}{1}{2q+1}{q}_q$ of Proposition~\ref{thm:floorfam} at
$n=2q+2$. For $q=2$ and $\sigma=2$ they are $\ea{13}{1}{11}{6}_2$ codes,
and a code with these parameters is among the code records released
with~\cite{BaekWonKim2026}. The code $\ea{9}{1}{7}{2}_2$ of
Table~\ref{tab:s-gen} attains the bound of Corollary~\ref{cor:k1} at
$\sigma=2$ with four fewer qudits and four fewer pairs.

For this code $\dim\Cd=n+k-c=8$ and $\dim\Hull=n-k-c=6$, so a basis of
$\Cd$ over $\F_2$ consists of six generators $S_1,\dots,S_6$ of $\Hull$
and two further operators $\bar X$ and $\bar Z$. Table~\ref{tab:s-gen}
gives such a basis. Each operator is written as a tensor product of Pauli
matrices and as the row $(x_1,\dots,x_9\,|\,z_1,\dots,z_9)$ of a binary
matrix. The Pauli matrix on
qudit $i$ is $X$ for $(x_i,z_i)=(1,0)$, $Z$ for $(0,1)$, $Y$ for $(1,1)$
and the identity, shown as a dot, for $(0,0)$.
\begin{table}[htbp]
\caption{Generators of the $\ea{9}{1}{7}{2}_2$ code as Pauli operators and as binary vectors.}
\label{tab:s-gen}
\centering
\setlength{\tabcolsep}{8pt}
\setlength{\arraycolsep}{3pt}
\delimitershortfall=-1pt \delimiterfactor=1000
\begin{tabular}{c|c}
\hline\hline
Pauli & $\vcenter{\kern5pt\hbox{$\left[\begin{array}{ccccccccc}
 X & \cdot & \cdot & \cdot & \cdot & \cdot & \cdot & \cdot & Y\\
 \cdot & Y & \cdot & \cdot & \cdot & X & \cdot & \cdot & \cdot\\
 \cdot & \cdot & Z & \cdot & \cdot & \cdot & \cdot & Z & \cdot\\
 \cdot & \cdot & \cdot & Z & \cdot & \cdot & \cdot & \cdot & Y\\
 \cdot & \cdot & \cdot & \cdot & X & X & \cdot & \cdot & \cdot\\
 \cdot & \cdot & \cdot & \cdot & \cdot & \cdot & Z & Z & \cdot\\
 \hline
 Y & \cdot & X & X & \cdot & X & X & Y & X\\
 \cdot & X & X & \cdot & Z & Y & Y & Y & Y\\
\end{array}\right]
\begin{array}{@{}l@{}}
\left.\vphantom{\begin{array}{c}0\\0\\0\\0\\0\\0\end{array}}\right\}\,S_1,\dots,S_6\\
\left.\vphantom{\begin{array}{c}0\\0\end{array}}\right\}\,\bar X,\bar Z
\end{array}$}\kern5pt}$\\
\hline
Binary & $\vcenter{\kern5pt\hbox{$\left[\begin{array}{ccccccccc|ccccccccc}
 1 & 0 & 0 & 0 & 0 & 0 & 0 & 0 & 1 & 0 & 0 & 0 & 0 & 0 & 0 & 0 & 0 & 1\\
 0 & 1 & 0 & 0 & 0 & 1 & 0 & 0 & 0 & 0 & 1 & 0 & 0 & 0 & 0 & 0 & 0 & 0\\
 0 & 0 & 0 & 0 & 0 & 0 & 0 & 0 & 0 & 0 & 0 & 1 & 0 & 0 & 0 & 0 & 1 & 0\\
 0 & 0 & 0 & 0 & 0 & 0 & 0 & 0 & 1 & 0 & 0 & 0 & 1 & 0 & 0 & 0 & 0 & 1\\
 0 & 0 & 0 & 0 & 1 & 1 & 0 & 0 & 0 & 0 & 0 & 0 & 0 & 0 & 0 & 0 & 0 & 0\\
 0 & 0 & 0 & 0 & 0 & 0 & 0 & 0 & 0 & 0 & 0 & 0 & 0 & 0 & 0 & 1 & 1 & 0\\
 \hline
 1 & 0 & 1 & 1 & 0 & 1 & 1 & 1 & 1 & 1 & 0 & 0 & 0 & 0 & 0 & 0 & 1 & 0\\
 0 & 1 & 1 & 0 & 0 & 1 & 1 & 1 & 1 & 0 & 0 & 0 & 0 & 1 & 1 & 1 & 1 & 1\\
\end{array}\right]
\begin{array}{@{}l@{}}
\left.\vphantom{\begin{array}{c}0\\0\\0\\0\\0\\0\end{array}}\right\}\,S_1,\dots,S_6\\
\left.\vphantom{\begin{array}{c}0\\0\end{array}}\right\}\,\bar X,\bar Z
\end{array}$}\kern5pt}$\\
\hline\hline
\end{tabular}
\end{table}
The operators $S_1,\dots,S_6$ commute with all eight operators of the
table and span
$\Hull$, and $\bar X$ and $\bar Z$ anticommute. The parameters were
re-derived from these generators alone by our evaluator and independently
in Magma. They are $k=1$, $c=2$ and $d=7$, and
$\bar X$ and $\bar Z$ have weight $7$. Here $\sigma=2$ and
$\Delta=2=(q-1)\sigma$, so the code attains the bound of
Corollary~\ref{cor:k1}. With Corollary~\ref{cor:k1} this gives
$\cmins(9,1,7)=2$. A code with the same parameters is among the code
records released with~\cite{BaekWonKim2026}.

\section{Machine checks and formalization}\label{sec:machine}

Every code reported here was evaluated exactly. The refutations behind
Proposition~\ref{prop:k2} were produced by CaDiCaL~3.0.0~\cite{CaDiCaL}.

\subsection{Evaluation of the codes}

Our evaluator takes a basis of $\Cd$ and computes the symplectic Gram
matrix, its rank, the hull and the least weight of a vector of
$\Cd\setminus\Hull$. For the entries marked exhaustive in
Tables~\ref{tab:codes} and~\ref{tab:fullrange} it enumerates all of $\Cd$.
The entries marked class by class concern the codes of
Eq.~\eqref{eq:diagcode}, whose hull is spanned by operators $\rho_s$ on
disjoint pairs. For them the evaluator enumerates the logical classes and
finds the least weight in each class pair by pair, by
Lemma~\ref{lem:pair}. The two modes agree wherever both run. The codes of
Eq.~\eqref{eq:diagcode} were evaluated with this evaluator for all $(q,k)$
with $q\le13$ prime and $k\le3$, and for $q=4$.

For the larger codes of Tables~\ref{tab:codes} and~\ref{tab:fullrange}
the class-by-class mode was rewritten in C++. It enumerates the logical
classes up to a nonzero scalar factor, at about $6\times10^7$ classes per
second on $14$ threads, with $1.1\times10^{10}$ classes for
$\ea{56}{4}{49}{24}_{27}$. It reproduces the distances of the codes with
$k=2$ obtained by the first evaluator and the distances that Magma V2.28
computes over $\F_8$ and $\F_9$. With the closed-formula scalars the Gram
matrix of the $2k$ logical generators has rank $2k$ for every prime
$q\le31$. In Magma, for $q\in\{8,9,16,25,27\}$, its rank falls short of
$2k$ exactly when the condition of Theorem~\ref{thm:diagcodes} fails. Magma
also computed the distance for $k\le2$, and for $k=3$ at $q=8,9$. The
codes of Table~\ref{tab:codes} found with a SAT solver were evaluated over
all of $\Cd$.

The entries marked G were computed with a separate implementation of
finite-field arithmetic, for all $71$ cases $(q,k)$ with $q\le13$ and
$1\le k\le q+1$ in Table~\ref{tab:fullrange}. In $24$ small cases the
distance was also evaluated class by class, and the other cases rest on
the lower bound. The code $\ea{8}{2}{5}{2}_3$ was also checked over all
$3^8$ vectors of $\Cd$. Magma re-derived the parameters of
$\ea{9}{1}{7}{2}_2$, $\ea{10}{2}{7}{3}_4$, $\ea{8}{1}{7}{3}_3$ and
$\ea{16}{1}{14}{7}_3$ from a basis of $\Cd$ alone, by full enumeration.

For $q=32$ and $q=64$ the entries marked G cover every $1\le k\le q/2$.
There we evaluated the codes of Eq.~\eqref{eq:diagcode} with the
closed-formula $\eta_s$ and with the first $\delta$, in the integer order of
Appendix~\ref{app:codes-ex}, that gives the Gram matrix full rank. The Gram matrix of the $2k$ logical generators has
rank $2k$, and the logical class of $\prod_{i=1}^{2k-1}(x-s_i)$, for
distinct nonzero $s_i$, contains a vector of weight $2q+3-2k$. With the
lower bound on the weight in the proof of Theorem~\ref{thm:diagcodes} this
gives the distance at every $k$.

\subsection{Refutations by satisfiability}

Each instance for Proposition~\ref{prop:k2} is the satisfiability problem
of its proof in Appendix~\ref{sec:length} for one value of
$|\mathrm{supp}\,\Hull|$, with $g_t$ and $f_t$ as in
Table~\ref{tab:appnotation} and $\mu_t$ as in the proof of
Theorem~\ref{thm:k1length}. Besides the $g_t$ and $f_t$, its unknowns are
the values of the $\mu_t$ on a complement $W$ of $\Hull$ in $\Cd$, the
common kernel of the $\mu_t$ of the five qudits of $T$. For the other
qudits the first nonzero coordinate of $f_t$ is $1$, and the $f_t$ are in
lexicographic order. The clauses express Eq.~\eqref{eq:radid}, the
conditions on the $f_t$ in the proof of Proposition~\ref{prop:k2}, and the
condition that every vector of $\Cd\setminus\Hull$ has at most $n-d=3$
zero components in $\mathrm{supp}\,\Hull$. CaDiCaL decides each instance in
seconds.

Three tests check that the encoding excludes only what it should. For
each of the $32$ choices of the set $T$, the two $\ea{10}{2}{7}{3}_4$ codes
we know, which have $k=\sigma=2$ and $\Delta=1$, can be brought to the form
above and then satisfy every clause. Configurations with a known
largest number of zero components, inserted into the encoding, are
accepted when the bound equals that number and rejected when it is one
less. In the analogous encoding for $k=1$ the fixed qudits carry all four
points of $\mathrm{PG}(1,3)$, and it finds the code $\ea{8}{1}{7}{3}_3$.

The same questions are also decided by a second encoding. It fixes the
$g_t$ and $\mu_t$ of the five qudits but not their $f_t$, and it imposes
only Eq.~\eqref{eq:radid} and the bound on zero components. For $q=2$ this
encoding and an exhaustive enumeration find no code. For $q=3$ it finds
none for $9\le|\mathrm{supp}\,\Hull|\le21$, except $16$ and $18$, which
were not run to completion.

For $k=1$ we also refuted directly the parameters with $q\le3$ and
$n\le13$ that Theorems~\ref{thm:diag}, \ref{thm:above}
and~\ref{thm:k1length} and Corollary~\ref{cor:k1} exclude, among them
$\ea{9}{1}{7}{1}_2$. Two encodings of the code itself are used. One fixes
a hull generator with only $z$-components, and the other takes a basis of
the hull in reduced row echelon form. Both find every instance
unsatisfiable, and with the target distance lowered by one each finds a
code, as it should.

Like those of Proposition~\ref{prop:k2}, the refutations of the instances
with $k=1$ and $q=2$ excluded by Theorem~\ref{thm:k1length} and of
$\ea{9}{1}{7}{1}_2$ come with DRAT proofs checked by \texttt{drat-trim}.
For the other instances with $k=1$ we report the decision of the solver.
The data record~\cite{DataRecord} also contains the two cross-checks of
Proposition~\ref{prop:k2}, the exhaustive enumeration for $q=2$ and the
second encoding, with DRAT proofs for $q=2$ and with instances and solver
logs for $q=3$.

\subsection{Comparison with codetables.de}

The tables of best-known codes at codetables.de~\cite{codetables}, checked
on 15 October 2026, cover $q=2$ and $q=3$. For each listed $(n,k,c)$ they
give a lower bound $d_L$ on the largest distance, attained by a known
code, and an upper bound $d_U$, and an entry is open when $d_L<d_U$. The
codes with $k=1$ and $c=n-3$ of Proposition~\ref{thm:unified} for $q=2$
and of Appendix~\ref{app:k1ex} for $q=3$ attain $d_U=n-1$ in fifty-four open
entries. These are all twenty-nine open entries with $k=1$ and $c=n-3$ of
the qubit table, at odd $7\le n\le63$, and all twenty-five such entries of
the qutrit table, at $12\le n\le36$. The same fifty-four entries are
reported in~\cite[Sec.~4.2]{BaekWonKim2026}. The qutrit code $\ea{11}{1}{10}{8}_3$
of Appendix~\ref{app:k1ex} has parameters for which the qutrit table has no entry.

These tables list no code with the parameters $\ea{6}{2}{3}{1}_2$,
$\ea{7}{2}{4}{2}_2$, $\ea{9}{2}{5}{2}_2$ or $\ea{12}{2}{8}{5}_3$ of
Table~\ref{tab:codes}. The best listed codes with the same $n$ and $k$
and smaller $c$ have distance $2$, $3$, $4$ and $7$. Each of the four has
the largest distance for its $n$, $k$ and $c$ (Theorem~\ref{thm:bdh} or
Proposition~\ref{prop:k2}). The code $\ea{6}{2}{3}{1}_2$ is listed by Lai
and Brun~\cite{LaiBrun2013}, and a code $\ea{7}{2}{4}{2}_2$ is among the
code records released with~\cite{BaekWonKim2026}.

We compared the codes of this paper over $\F_2$ and $\F_3$ with the
entries of the same tables, including those whose construction the tables
describe as teleportation. Our codes are carried to more pairs by
Lemma~\ref{lem:lift}, which does not lower the distance, and only entries
that the tables list are compared. Two counts are kept apart. The first
counts the choices of $(q,n,k,d)$ for which one of our codes of distance
$d$ uses fewer pairs than the cheapest listed code of distance at least
$d$. The second counts the listed entries $(q,n,k,c)$ whose $d_L$ is
exceeded by one of our codes with at most $c$ pairs. Both counts are
repeated with the tables augmented by the published codes they do not
list. These are the binary $\ea{n}{1}{n-1}{n-3}_2$ codes of Li, Ren, Guan
and Liu~\cite{LiRenGuanLiu2026} for every $n\ge4$, the code
$\ea{6}{2}{3}{1}_2$ of Lai and Brun, and the qubit and qutrit codes
of~\cite{BaekWonKim2026}. These are the codes of
Proposition~\ref{thm:unified} and the $92$ code records described in
Sec.~4.1 of~\cite{BaekWonKim2026}. Table~\ref{tab:s-counts} gives both
counts.
\begin{table}[htbp]
\caption{Codes of this paper that improve on the tables of codetables.de.}
\label{tab:s-counts}
\centering
\setlength{\tabcolsep}{8pt}
\begin{tabular}{c|c|cc|cc}
\hline\hline
\multirow{2}{*}{Baseline} & \multirow{2}{*}{$q$} & \multicolumn{2}{c|}{Fewer pairs} & \multicolumn{2}{c}{Larger distance}\\
\cline{3-6}
 & & Cases & Largest & Entries & Reach $d_U$\\
\hline
\multirow{2}{*}{Tables} & $2$ & $73$ & $20$ & $82$ & $57$\\
 & $3$ & $35$ & $12$ & $74$ & $70$\\
\hline
\multirow{2}{*}{\begin{tabular}{@{}c@{}}Tables and\\published codes\end{tabular}} & $2$ & $35$ & $20$ & $50$ & $26$\\
 & $3$ & $26$ & $12$ & $24$ & $20$\\
\hline\hline
\end{tabular}
\end{table}
The largest gain in distance is $10$ for $q=2$ and $3$ for $q=3$. The
three arrows of Fig.~\ref{fig:map}, with $1$, $2$ and $6$ pairs fewer, are
among the first count, and no $d_U$ is exceeded. The two counts overlap in
the codes they use and are not added. Scripts in the data
record~\cite{DataRecord} reproduce both counts. They also evaluate
exactly the $133$ codes of Propositions~\ref{thm:floorfam},
\ref{thm:unified} and~\ref{prop:k1codes} in the range of the tables.

A second check runs the other way. None of the reported lower distances in
the $11{,}043$ entries with $k\ge1$ in the qubit and qutrit tables of
best-known EAQECCs~\cite{codetables}, as of 10 September 2026 and
unchanged on 15 October 2026, including the $552$
teleportation-marked entries, exceeds the upper bounds on $d$ given by
Theorems~\ref{thm:bdh} and~\ref{thm:diag}, or that of
Theorem~\ref{thm:above} with $N_q$ replaced by its Griesmer bound.

\subsection{Formalization}

The results are formalized in the Lean~4 proof assistant~\cite{Lean4},
version 4.29.0-rc6, with its mathematical library Mathlib~\cite{Mathlib} at
commit \texttt{1ea3462986d0}. Each result is stated in a separate Lean
file, a challenge, that contains only definitions and the statement. The
tool Comparator~\cite{Comparator} checks that a proof, kept in another
file, proves exactly this statement and uses only the standard axioms. A
result counts as formalized when Comparator accepts its proof and rejects a
control in which one statement of the same challenge has been altered.
With the range of each statement given below, this holds for every lemma,
theorem, proposition and corollary of this paper. For
Proposition~\ref{prop:k2} the reduction, the Boolean encodings and the
finite refutations are also checked by the Lean kernel. The cited ranges of
Fan~\cite{Fan2026} for odd $q$ and the range $\ell\le2p-2$, with their
duals, for $2\le\ell\le q$, are also formalized. Some cited $M_q$ values
and the numerical code evaluations in
Tables~\ref{tab:codes}--\ref{tab:fullrange} remain outside the
formalization. Figure~\ref{fig:deps} marks the status of each result.

We call a result certified when it is formalized in this sense, Comparator
rejects every prescribed control of its challenge, and the source files
used by its proof are unchanged since the Comparator run, as the hashes
recorded in the data record~\cite{DataRecord} show. The Lean definitions
follow Sec.~\ref{sec:prelim} for codes, qudit components and the
symplectic form, and Sec.~\ref{sec:further} for $N_q$ and $B_q$.

The certified results are the distance bounds, the constructions, the
identity~\eqref{eq:radid} and the bounds on the support of the hull shown
in Fig.~\ref{fig:deps}. In the Lean statements the distance is the least
weight of a vector of $\Cd\setminus\Hull$, as in Sec.~\ref{sec:prelim},
and each statement that a bound is attained asserts the existence of a
code that attains it. In Theorem~\ref{thm:diag}, $M_q(2k)$ is defined in
Lean as the largest length of an MDS code of dimension $2k$, not assumed.
Corollaries~\ref{cor:diagexact} and~\ref{cor:diag} are formalized with
$M_q(2k)=q+1$ as a hypothesis wherever they assume it.
Lemma~\ref{lem:peel} includes the shortened code, its symplectic dual and hull, with parameters $\ea{n-1}{k}{d'}{c+1}_q$ and distance $d'\ge d$.
The Lean statements concern subspaces of $\F_q^{2n}$ and imply the bounds
on codes as stated in this paper.

Besides the ranges of Fan and of $\ell\le2p-2$, the certified values of
$M_q(\ell)$ are $M_q(\ell)=\ell+1$ for $\ell\ge q$, $M_q(2)=q+1$,
$M_q(4)=M_q(q-2)=q+1$ for $q\ge4$, $M_q(q-1)=q+1$ for odd $q$ and, in
characteristic two, $M_q(5)=q+1$ for $q\ge8$, together with the duality of
MDS codes.

The formalization of Theorem~\ref{thm:s-full-range} covers every finite
field and every $1\le k\le q+1$, and the scalars of
Table~\ref{tab:s-scalars} and the two examples that follow the table are
also checked in Lean. Like the code evaluations, the comparisons with
codetables.de are not checked in Lean.

\end{document}